\documentclass[12pt]{aastex}
\usepackage{natbib}

\def\folio{\ifnum\pageno=1\nopagenumbers\else\number\pageno\fi}

\def\lax    {\ifmmode{_<\atop^{\sim}}\else{${_<\atop^{\sim}}$}\fi}
\def\gax    {\ifmmode{_>\atop^{\sim}}\else{${_>\atop^{\sim}}$}\fi}
\newbox\grsign      \setbox\grsign=\hbox{$>$}
\newdimen\grdimen   \grdimen=\ht\grsign
\newbox\simgreatbox \setbox\simgreatbox=\hbox{\raise.5ex\hbox{$>$}\llap
                        {\lower.5ex\hbox{$\sim$}}}\ht1=\grdimen\dp1=0pt
\newbox\simlessbox  \setbox\simlessbox =\hbox{\raise.5ex\hbox{$<$}\llap
                        {\lower.5ex\hbox{$\sim$}}}\ht2=\grdimen\dp2=0pt
\def\simgreat{\mathrel{\copy\simgreatbox}}

\def\pz {\phantom{$>$}}

\def\boxit#1    {\vbox{\hrule\hbox{\vrule\kern3pt
                  \vbox{\kern3pt#1\kern3pt}\kern3pt\vrule}\hrule}}
\def\h      {\ifmmode{^{\rm h}}\else{$^{\rm h}$}\fi}
\def\m      {\ifmmode{^{\rm m}}\else{$^{\rm m}$}\fi}
\def\s      {\ifmmode{^{\rm s}}\else{$^{\rm s}$}\fi}
\def\arc    {\ifmmode{{\rlap.}{''}}\else{${\rlap.}{''}$}\fi}
\def\as    {\ifmmode{{\rlap.}{''}}\else{${\rlap.}{''}$}\fi}
\def\arcmin    {\ifmmode{{\rlap.}{'}}\else{${\rlap.}{'}$}\fi}
\def\am    {\ifmmode{{\rlap.}{'}}\else{${\rlap.}{'}$}\fi}
\def\mum     {\ifmmode{\mu{\rm m}}\else{$\mu{\rm m}$}\fi}
\def\decdeg {\rlap . {}^\circ}     
\def\kms    {\ifmmode{{\rm km~s}^{-1}}\else{km~s$^{-1}$}\fi}

\def\ccm    {cm$^{-3}$}
\def\scm    {cm$^{-2}$}

\def\Lsun   {$L_{\odot}$}

\def\Msun   {$M_{\odot}$}
\def\Mspy   {\ifmmode {M_{\odot} {\rm yr}^{-1}} \else $M_{\odot}$~yr$^{-1}$\fi}
\def\Mdot   {\ifmmode {\dot M} \else $\dot M$\fi}
\def\mhd    {\ifmmode {n_{{\rm H}_2}} \else $n_{{\rm H}_2}$\fi}
\def\mhcd   {\ifmmode {N_{{\rm H}_2}} \else $N_{{\rm H}_2}$\fi}

\def\El      {\ifmmode{E_{\ell}}\else{$E_{\ell}$}\fi}
\def\beam    {\ifmmode{\theta_{\rm B}}\else{$\theta_{\rm B}$}\fi}
\def\Trot   {\ifmmode{T_{\rm rot}}\else$T_{\rm rot}$\fi}

\def\Teff   {\ifmmode{T_{\rm eff}}\else$T_{\rm eff}$\fi}

\def\TAS    {$T_{\rm A}^{*}$}

\def\Tkin   {$T_{\rm kin}$}

\def\ITRS   {\ifmmode{\smallint {\rm T}_{R}^{*}dv}\else{$\smallint
{\rm T}_{R}^{*}dv$}\fi}
\def\ITRS   {\ifmmode{\smallint {\rm T}_{R}^{*}dv}\else{$\smallint
{\rm T}_{R}^{*}dv$}\fi}
\def\ITAS   {\ifmmode{\smallint {\rm T}_{A}^{*}dv}\else{$\smallint
{\rm T}_{A}^{*}dv$}\fi}

\def\hh         {H$_2$}

\def\hhco       {H$_2$CO}

\def\hzo        {H$_2$O}

\def\nhhh       {NH$_3$}

\def\ceto       {C$^{18}$O}

\def\ccchh      {C$_3$H$_2$}
\def\meth   {CH$_3$OH}

\def\zkok   {\hbox{$2_k - 1_k$}}

\def\etrans             {\hbox{$2_{0} - 3_{-1}E$}}

\def\TSGH  {\hbox{$4_{-1} - 3_{0}E$}}
\def\EFGH  {\hbox{$5_{-1} - 4_{0}E$}}

\def\ohn  {\hbox{$0_{0} - 1_{-1}E$}}

\def\ffs   {\hbox{$1_{10} - 1_{01}$}}

\def\go     {G1.6$-$0.025}

\def\lefttitle#1  {\noindent \hangindent=18.0pt \hangafter=1 {#1} \par}
\def\apjl   {{\sl Ap.~J.\ (Letters)\ }}
\def\apjs   {{\sl Ap.~J.\ Suppl.\ }}

\def\vol#1  {{\bf {#1}{\rm,}\ }}
\font\tenssb=cmssbx10
\font\tenbf=cmbx10
\font\sevenbf=cmbx8
\font\fivebf=cmbx6
\def\doublespace {\smallskipamount=6pt plus2pt minus2pt
                  \medskipamount=12pt plus4pt minus4pt
                  \bigskipamount=24pt plus8pt minus8pt
                  \normalbaselineskip=24pt plus0pt minus0pt
                  \normallineskip=2pt
                  \normallineskiplimit=0pt
                  \jot=6pt
                  {\def\smallskip {\vskip\smallskipamount}}
                  {\def\medskip   {\vskip\medskipamount}}
                  {\def\bigskip   {\vskip\bigskipamount}}
                  {\setbox\strutbox=\hbox{\vrule
                    height17.0pt depth7.0pt width 0pt}}
                  \parskip 12.0pt
                  \normalbaselines}
\def\unetdemi    {\smallskipamount=6pt plus2pt minus2pt
                  \medskipamount=12pt plus4pt minus4pt
                  \bigskipamount=24pt plus8pt minus8pt
                  \normalbaselineskip=16pt plus0pt minus0pt
                  \normallineskip=2pt
                  \normallineskiplimit=0pt
                  \jot=6pt
                  {\def\smallskip {\vskip\smallskipamount}}
                  {\def\medskip   {\vskip\medskipamount}}
                  {\def\bigskip   {\vskip\bigskipamount}}
                  {\setbox\strutbox=\hbox{\vrule
                    height17.0pt depth7.0pt width 0pt}}
                  \parskip 12.0pt
                  \normalbaselines}
\def\singlespace {\smallskipamount=3pt plus1pt minus1pt
                  \medskipamount=6pt plus2pt minus2pt
                  \bigskipamount=12pt plus4pt minus4pt
                  \normalbaselineskip=12pt plus0pt minus0pt
                  \normallineskip=1pt
                  \normallineskiplimit=0pt
                  \jot=3pt
                  {\def\smallskip {\vskip\smallskipamount}}
                  {\def\medskip   {\vskip\medskipamount}}
                  {\def\bigskip   {\vskip\bigskipamount}}
                  {\setbox\strutbox=\hbox{\vrule
                    height8.5pt depth3.5pt width 0pt}}
                  \parskip 4pt
                  \normalbaselines}
\def\smallerspace {\smallskipamount=3pt plus0pt minus0pt
                  \medskipamount=6pt plus0pt minus0pt
                  \bigskipamount=10.5pt plus0pt minus0pt
                  \normalbaselineskip=10.5pt plus0pt minus0pt
                  \normallineskip=1pt
                  \normallineskiplimit=0pt
                  \jot=3pt
                  {\def\smallskip {\vskip\smallskipamount}}
                  {\def\medskip   {\vskip\medskipamount}}
                  {\def\bigskip   {\vskip\bigskipamount}}
                  {\setbox\strutbox=\hbox{\vrule
                    height8.5pt depth3.5pt width 0pt}}
                  \parskip 0pt
                  \normalbaselines}
\def\memospace    {\smallskipamount=4pt plus1pt minus1pt
                  \medskipamount=6pt plus2pt minus2pt
                  \bigskipamount=14pt plus6pt minus6pt
                  \normalbaselineskip=14pt plus0pt minus0pt
                  \normallineskip=1pt
                  \normallineskiplimit=0pt
                  \jot=4pt
                  {\def\smallskip {\vskip\smallskipamount}}
                  {\def\medskip   {\vskip\medskipamount}}
                  {\def\bigskip   {\vskip\bigskipamount}}
                  {\setbox\strutbox=\hbox{\vrule
                    height17.0pt depth7.0pt width 0pt}}
                  \parskip 2.0pt
                  \normalbaselines}
\def\memowidespace    {\smallskipamount=5pt plus1pt minus1pt
                  \medskipamount=7.5pt plus2pt minus2pt
                  \bigskipamount=17.5pt plus6pt minus6pt
                  \normalbaselineskip=17.0pt plus0pt minus0pt
                  \normallineskip=1.25pt
                  \normallineskiplimit=0pt
                  \jot=5pt
                  {\def\smallskip {\vskip\smallskipamount}}
                  {\def\medskip   {\vskip\medskipamount}}
                  {\def\bigskip   {\vskip\bigskipamount}}
                  {\setbox\strutbox=\hbox{\vrule
                    height21.25pt depth8.75pt width 0pt}}
                  \parskip 2.5pt
                  \normalbaselines}
\newbox\grsign \setbox\grsign=\hbox{$>$} \newdimen\grdimen \grdimen=\ht\grsign
\newbox\laxbox \newbox\gaxbox
\setbox\gaxbox=\hbox{\raise.5ex\hbox{$>$}\llap
     {\lower.5ex\hbox{$\sim$}}}\ht1=\grdimen\dp1=0pt
\setbox\laxbox=\hbox{\raise.5ex\hbox{$<$}\llap
     {\lower.5ex\hbox{$\sim$}}}\ht2=\grdimen\dp2=0pt
\def\gax{\mathrel{\copy\gaxbox}}
\def\lax{\mathrel{\copy\laxbox}}
\def\deg     {\ifmmode {\rlap.}$\,$^\circ$\,$\! \else ${\rlap.}$\,$^\circ$\,$\!$\fi}
\def\as     {\ifmmode {\rlap.}$\,$''$\,$\! \else ${\rlap.}$\,$''$\,$\!$\fi}
\def\s      {\ifmmode {\rlap.}$\,$^{s}$\,$\! \else ${\rlap.}$\,$^{s}$\,$\!$\fi}
\def\kms    {\ifmmode{{\rm ~km~s}^{-1}}\else{~km~s$^{-1}$}\fi}

\def\Msun   {$M_{\odot}$}
\def\apjs   {{ApJS}}

\def\aap    {{A\&A}}

\begin{document}
\title{MOLECULES IN G1.6$-$0.025 -- ``HOT'' CHEMISTRY
IN THE ABSENCE OF STAR FORMATION AT THE PERIPHERY OF THE GALACTIC CENTER REGION}
\author{Karl M. Menten\altaffilmark{1}, Robert W. Wilson\altaffilmark{2},
Silvia Leurini\altaffilmark{3}, and Peter Schilke\altaffilmark{1}}
\altaffiltext{1}{Max-Planck-Institut f\"ur Radioastronomie, Auf dem H\"ugel,
D-53121 Bonn, Germany\email{kmenten, pschilke@mpifr-bonn.mpg.de}}
\altaffiltext{2}{Harvard-Smithsonian Center for Astrophysics, 60 Garden
Street/MS42, Cambridge MA 02138\email{rwilson@cfa.harvard.edu}}
\altaffiltext{3}{European Southern Observatory, Karl-Schwarzschild-Strasse 2,
D-85748 Garching\email{sleurini@eso.org}}

\begin{abstract}
We present molecular line mapping of the Giant Molecular Cloud G1.6$-$0.025,
which is located at the high longitude end of the Central Molecular
Zone of our Galaxy. We assess the degree of star formation activity
in that region using several tracers and find very little.
We made a large scale, medium ($2'$) resolution map in the $J = 2 - 1$ transition of SiO
for which we find clumpy emission over a $\sim 0\decdeg8\times0\decdeg3$-sized region
stretching along the Galactic plane. Toward selected positions we also took spectra in the
easy to excite $J_k=2_k-1_k$ quartet of \meth\ and the CS $2 - 1$ line. Throughout the cloud
these \meth\ lines are, remarkably, several times stronger than, both, the CS and the SiO lines.
The large widths of all the observed lines, similar to values generally found in the
Galactic center, indicate a high degree of turbulence. Several high LSR velocity clumps
that have 50--80 \kms\ higher velocities than the bulk of the molecular cloud appear at
the same projected position as ``normal'' velocity material; this may indicate cloud-cloud
collisions. Statistical equilibrium modeling of the
\meth\ lines observed by us and others yield relatively high densities and moderate
temperatures for a representative dual velocity position. 
We find $8~10^4$ cm$^{-3}$/30 K for material in the G1.6$-$0.025 cloud and a higher
temperature (190 K), but a 50\%\ lower density in a high velocity clump projected on
the same location. Several scenarios are discussed in which shock chemistry might enhance
the \meth\ and SiO abundances in G1.6$-$0.025 and elsewhere in the Central Molecular Zone.
\end{abstract}

%
\section{\label{intro}INTRODUCTION -- THE CENTRAL MOLECULAR ZONE AND ITS (LITTLE) STAR FORMATION ACTIVITY}
``Ordinary'' giant molecular clouds (GMCs) in the Galactic disk
have sizes of tens
of parsecs, temperatures, \Tkin, in the low tens of K, and densities, $n$, of
order a hundred cm$^{-3}$. Embedded in these GMCs are regions
of star formation with much higher temperatures and densities.
In contrast, the gas in the central molecular
zone \citep[CMZ; see e.g.\ ][]{MorrisSerabyn1996} of our Galaxy,
stretching from Galactic latitude,$l$, $\sim+1\decdeg6$  to $-1\decdeg1$ in a $\sim\pm
0\decdeg3$ wide band in latitude, $b$, around the Galactic center \citep[GC, ][]{Bally_etal1987,Bally_etal1988,Dahmen_etal1997,Dahmen_etal1998,Oka_etal1998}
is in general characterized by much higher temperatures, densities, 
and more turbulence, resulting in larger linewidths ($\simgreat 10$--30 \kms). \citet{GuestenPhilipp2004} give  a recent review.

To get a handle on the chemistry and physical conditions
in these peculiar clouds, multi-transition measurements of molecules
other than the ubiquitous and easily-thermalized carbon monoxide (CO) are highly desirable.
By observing many lines from a given species and modeling
the results using, e.g., large velocity gradient (LVG)
methods, one can derive the densities and temperatures of the GC clouds,
which are known to be significantly higher than values in Galactic disk clouds,
but still relatively ill-constrained by existing data. Extensive surveys of
carbon monosulfide (CS), $^{13}$CO, and \ceto\ have been made mostly  with
coarse spatial resolutions of $2'$ and $9'$, respectively
\citep{Bally_etal1987,Dahmen_etal1997} and, in a smaller region around the GC
itself, of cyanoacetylene (HC$_3$N), ammonia (\nhhh), and silicon monoxide
(SiO) with beam sizes between $40''$ and $140''$
\citep{Guesten_etal1981, Walmsley_etal1986, Martin-Pintado_etal1997}.
Walmsley et al. observed
several HC$_3$N transitions within $1{\rlap.}{'}5$ of the GC
and found that the bulk of the gas has $T \approx 80$ K
and $n \approx 10^4$ \ccm, while 20\%\ of its mass may exist
in higher ($10^5$ \ccm) density clumps.
Astonishingly, a large number of complex organic molecules have been found to show widespread emission all over the CMZ (see \S\ref{cms}).

Whether tidal forces, stronger turbulence, higher densities, and
stronger magnetic fields (compared to Galactic disk clouds) favor
stars formation or hinder it can be argued either way. Fact is that
the spectacular Arches and Quintuplet clusters give testimony
for violent star formation a few million years ago \citep{Figer_etal1999,Figer_etal2002}.
\citet{Figer_etal2004} argue that ongoing star formation is responsible
for the observed central stellar cusp.
From Infrared Space Observatory (ISO) and Spitzer Space Telescope infrared wavelength data \citet{Schuller_etal2005} derive
a star formation rate in the CMZ of 0.2 \Mspy\ over the past $\sim 0.5$ Myr,
which is an appreciable fraction of our whole Galaxy's star formation rate.

However, apart from the star formation-wise extremely active Sgr B2 region,
little of what is commonly assumed to be evidence for \textit{on-going} high-mass
star formation is found in
the GC GMCs at present, such as (ultra)compact HII regions, \hzo\ and \meth\
masers.
The submillimeter-detected dust ridge identified by \citet{LisCarlstrom1994} marks highest column density material of the general GC dust emission/molecular cloud distribution \citep{Pierce-Price_etal2000}. It has the potential to
harbor proto-cluster cores with the massive ($1~10^6$ \Msun)
Infrared Dark Cloud M0.25+0.11 the most prominent
example \citep{Lis_etal1994,LisMenten1998}.
However, in this ridge, namely in M0.25+0.11,  and also in the larger GC region
only few markers of star formation have been found, just very few \hzo\ masers and
compact continuum sources \citep{GuestenDownes1981,Lis_etal1994}.
\citet{Caswell1996} covered an area of extent  $l = \pm 0\as9, b = \pm 0\as5$
in a sensitive interferometric search for 6.7 GHz class II methanol masers, which
are unambiguous tracers of \textit{high-mass} star formation.
He only found 23 masers sites with $\sim$\textit{half} of them located in the Sgr B2 complex alone.


\section{\label{peculiar}THE PECULIAR \go\ MOLECULAR CLOUD}
\subsection{\label{ewmg}Extended, Warm, Molecular Gas}
\begin{figure}
\includegraphics[width=16cm]{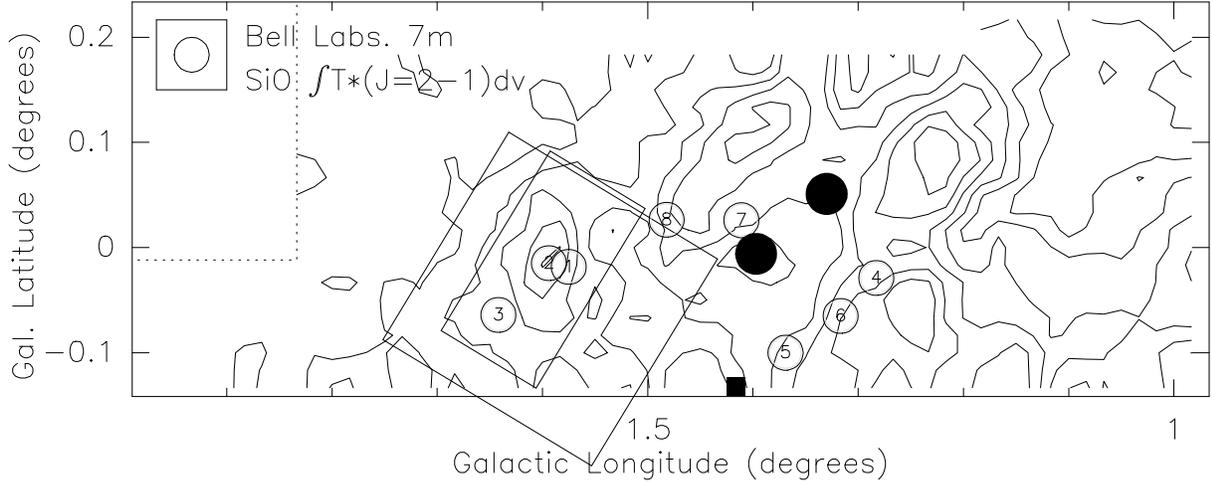}
\caption{
Map of integrated SiO $J=2-1$ emission of G1.6$-$0.025 made with the Bell Labs.
7m telescope. Contours
are 2 to 12  in steps of 2 times 1.8 K\kms\, which is equal to the
rms noise. The area within the
\textit{dotted line} was not mapped.
The $2'$ diameter beam (FWHM)  is indicated in the left upper
corner of the \textit{left upper panel}. Fiducial positions lying within the
map boundaries are indicated. The square gives the extent of
the ammonia map shown in Fig. 3 of Gardner et al. 1985 and the rectangle
within it the area mapped by Salii et al. 2002 in methanol lines.
Methanol absorption in the \etrans\ line was mapped by
Whiteoak \&\ Peng 1989  over a roughly similar area as ammonia.
The \ccchh\ spectra presented by Kuiper et al. 1993 were taken
toward various locations within that area. The two dots mark the
positions of the radio/FIR sources Sgr D7 (western) and D8
(eastern source). The square marks the position of a 1720 MHz OH maser.
}\label{fig:integratedmap}
\end{figure}

The molecular cloud \go\footnote{Strictly speaking \go\ is
only the cloud fragment delineated by the square in Fig.
\ref{fig:integratedmap}.
For simplicity's sake, we refer by that name to the
whole  $\sim 0\decdeg8 \times 0\decdeg35$-sized region shown
in that figure.}
lies at the very  easternmost longitude edge of
the cloud complexes making up the CMZ that were mapped
in CO, $^{13}$CO, \ceto\, CS, and other
molecules \citep{Bally_etal1987,Bally_etal1988,Dahmen_etal1997,Dahmen_etal1998, Oka_etal1998,Martin_etal2004} Maybe because of its location, this cloud received
comparatively little attention in the past. However, the few
observations that do exist reveal a fascinating chemical picture.

For further reference, we show, in Fig. \ref{fig:integratedmap}, an image of the velocity-integrated emission in the $J = 2 - 1$ transition of SiO discussed in \S \ref{MED}.

Apart from the low resolution surveys in CS and \ceto\,
\go\ has been mapped in several
inversion transitions of \nhhh\ by \citet{Gardner_etal1985},
who found remarkably strong emission in the $(J,K)$ =(1,1), (2,2),
and (3,3) lines, with the (3,3) line possibly being
inverted toward one position. The \nhhh\ observations indicate
that the cloud is warm (probably $> 50$ K) but Gardner et al.
do not give a thorough discussion of the temperatures suggested by
the \nhhh\ data. \citet{GardnerBoes1987}, based
on another set of \nhhh\ data, conclude $T> 120$ K.
\cite{Kuiper_etal1993} observed the \ffs\
transition of \ccchh\ near 18.3 GHz. This line, whose appearance
is ubiquitous in molecular clouds, here appears atypically
in {\it enhanced absorption} against the cosmic microwave background
radiation.

Also overcooling was found by \citet{WhiteoakPeng1989} in the
\etrans\ line of \meth. This transition, which is the
second strongest class II methanol {\it maser} line\footnote{See \citet{Menten1991}
for the nomenclature/classification of \meth\ masers.}
\citep{Batrla_etal1987}, has been
found in absorption in dark clouds \citep{Walmsley_etal1988}, which lack the far-infrared
continuum emission necessary to produce class II maser action
and also in absorption in class I methanol maser regions for which the same is true.
\citet{WhiteoakPeng1989} model the very (up to 2 K) deep
absorption and find a high  $E$-type \meth\ abundance
of $10^{-7}$ and an \hh\ density $\sim 10^4$ \ccm.
\citet{HaschickBaan1993} find {\it extended}, strong {\it maser}
emission in the (class I) methanol  maser \TSGH\ line. As
explained in \S \ref{metstat}, the occurrence of
overcooling in the \etrans\ and simultaneous maser
action in the \TSGH\ lines is expected from basic properties
of the methanol molecule and can be used to constrain
the physical conditions in the masing region.

\citet{Sobolev1992} and \citet{Salii_etal2002}
discuss observations of several
lines from methanol and other molecules
with the RT-22 telescope at Pushchino (with $2'$ resolution) and the Swedish-ESO Submillimeter Telescope (SEST) (with $\approx0\am5$ and $\approx1'$  resolution), which they model to derive physical parameters. Their results
are compared to our own in \S \ref{metstat}.

One of the most astonishing things about \go\ is the virtually complete
absence of any marker of high-mass star formation on which
we shall elaborate in \S \ref{nosf}.
This means, in \go\ we have the opportunity to study large-scale
high-temperature gas-phase chemistry in the absence of strong UV fields,
which, in star-forming regions occurs only in very high extinction,
spatially compact
hot cores near young (proto)stars. In the latter environment, the high
observed abundances of SiO, \meth, and other more complex species, which are
orders of magnitudes or more higher than
their quiescent, cool
molecular cloud abundances, are thought to result from the evaporation
of ice grain mantles in which these molecules resided in
frozen-out form \citep[see, e.g.\ ][and references therein]{GarrodHerbst2006}.  Later in this paper (in \S \ref{shockchemistry})
we will argue that in the
case of \go\ these abundances may be the result of shock chemistry.

Given that several of the cloud's characteristics (high linewidths,
high abundances of usually rare species) are
typical for GC GMCs, we assume in the following that is at the
distance of the GC, 8 kpc \citep{Reid1993} . Other evidence for placing the cloud
there are the high [H$_2^{13}$CO]/[\hhco] and
[H$_2^{13}$CO]/[H$_2$C$^{18}$O]  isotopic abundance
ratios, which are by factors of  4 and 2 higher, respectively, than values
found in the solar neighborhood, but typical for the GC region
\citep{GardnerWhiteoak1981,GardnerWhiteoak1982}.
Its projected distance from the GC is ca. 200 pc.

In this paper we report medium spatial resolution ($2'$) mapping
observations of \go\ in the $J = 2-1$ rotational line of
silicon monoxide (SiO). Furthermore, we observed selected
positions in the $2-1$ line of carbon monosulfide (CS) and several
methanol (\meth) transitions. As described by \citet{Leurini_etal2004}, newly
calculated collisional rate coefficients now allow meaningful modeling
of methanol excitation to obtain densities and kinetic
temperatures (see \S \ref{metstat}).  Using all this information, we summarize the available chemical
information for \go\ and in \S\ref{shockchemistry} discuss possible reasons for the
observed picture.

\subsection{(Almost Non-)Existent Active Massive Star Formation
in \go}\label{nosf}
In order to assess whether \go's peculiar chemistry could be energetically driven by the influence of young high-mass stars, we conducted  a census of observational phenomena and, in particular, of
tracers of on-going high-mass star formation activity in the region and its
surroundings. For this, we conducted a literature search using the
SIMBAD\footnote{http://simbad.u-strasbg.fr/Simbad} database.
We searched for all astronomical objects contained in that
database in a circle of $0\decdeg3$ radius with
$l,b = 1\decdeg4,0\decdeg0$
at its center. We found a total of 138 objects, many of which are
foreground (some of them OH/IR) stars and planetary nebulae.
However, also a few compact radio continuum sources were found, for some of which,
as discussed in the following, multi-wavelength data
are available,  allowing a characterization.

For our purposes, we are only interested in objects that
are (most probably) associated with \go\ and may have
some influence on their environment. Given this, it
is straightforward to eliminate radio sources from our list that
are not coincident with IRAS point sources. A spot check
reveals that, e.g., G1.285$-$0.054, which has no associated
IRAS source is, both, variable and has a negative, non-thermal spectral
index and is therefore in all likelihood of extragalactic
origin. On the other hand, IRAS sources without detected
radio emission may be important, as there is {\it no}
established correlation between the radio and IR luminosities
of very young high-mass protostellar objects; see the famous
case of Orion IRc2 \citep{MentenReid1995}.

OH/IR objects
are easy to eliminate from our list since they can be identified
by their IRAS colors \citep[as established by ][]{vanderVeenHabing1988}
and/or the presence of 1612 MHz OH maser
emission.

The  only IRAS sources in the general region that are definitely associated
with star formation are IRAS 17450$-$2742 and  17457$-$2743\, which are
coincident  with the compact HII regions Sgr D 7 and 8, respectively
\citep{Liszt1992}; the latter one is also known as GPSR5 1.396-0.006
\citep{Becker_etal1994}. We note that these sources are \textit{not} coincident
with any molecular peak in the cloud. Observed and derived properties for them
can be found in Table \ref{d7d8}. While Liszt determines for Sgr D 7 a size of
$1'$ at 1.6 GHz, he finds source 8 unresolved in his $13'' \times 23''$ beam.
Becker et al. (2004), with $4''$ beam size, do not detect 7 at either 1.4 or 5
GHz, most certainly ``resolving it out'' and find source 8 unresolved at either
frequency with flux densities of 24.2 and 26 mJy at 1.4 and 5 GHz,
respectively. Assuming a size of $2''$ for source 8 and an electron temperature
of 10000 K
we calculate an optical depth, $\tau$,
of the free-free emission of 0.47, 0.29, and 0.03     at 1.4, 1.6,  and 5 GHz
respectively. For
source 7 we determine $\tau = 0.003$.
Using the formula given by \citet{Mezger_etal1974} we derive that Lyman continuum photon fluxes
of $1.1\times10^{48}$ s$^{-1}$ and $1.4\times10^{47}$ s$^{-1}$ are needed to produce
the compact HII regions Sgr D 7 and 8, respectively.
According to the  Tables given
by \citet{Panagia1973}
these values correspond to ZAMS spectral types of O9 and B0, respectively.

To determine the neutral gas masses and luminosities of these sources
from the IRAS data in the same way as described by  \citet{LisMenten1998},  Planck functions were fitted to the
measured 12, 25, 60, and 100 $\mu$m flux densities\footnote
{This fitting was not straightforward as the 12, 60, and 100
$\mu$m flux densities listed by the SIMBAD database for Sgr D 7
are very uncertain, as is  the 60 $\mu$m value for
D 8; for the latter source only upper limits are given for
the 12  and 100 $\mu$m flux densities.} to determine
dust temperatures
(see Table \ref{d7d8}).
We used the formulae given therein and
in the paper by \citet{Motte_etal2003} to
determine  the gas masses listed in Table \ref{d7d8}
for  Sgr D 7 and 8 (taking the upper
limits in the table at face value).
Integrating over the
spectral energy
distributions we derive the  bolometric luminosities given
in Table \ref{d7d8}, which, again according to \citet{Panagia1973}, are produced by a
B0 and a
B1 ZAMS star, respectively.
These spectral types are very similar to the ones inferred from the
Lyman continuum fluxes.

All in all we conclude that, apart from the two sources discussed above, there is presently very little star-forming
activity  in \go\ and consider it highly
unlikely that star-formation activity contributes to its
enhanced temperature  in a significant way.

\subsection{A Possible Connection to the G1.4$-$0.1 Supernova Remnant}\label{SNR}
The supernova remnant (SNR) G1.4$-$0.1, which lies partially in
the area mapped by us clearly interacts with molecular gas.
\citet{Yusef-Zadeh_etal1999} found 1720 MHz OH maser emission
at  $l,b = 1\decdeg4164, -0\decdeg1323$ (see Fig. \ref{fig:integratedmap})
at a velocity of
$-2.4$ \kms, which is blue-shifted relative to almost all
of the gas we observe. Possibly the masing gas is located
in the portion of the swept-up molecular material coming
toward us and thus amplifying the continuum background.
\citet{Lockett_etal1999} constrain the conditions
for the occurrence of these masers to temperatures between 50
and 125 K, densities and OH column densities around
$10^5$ \ccm\ and $10^{16}$ \scm, respectively, typical for post magneto-hydrodynamic (``C'')-shock material.
These numbers can be compared with the values derived from our methanol
modeling in \S \ref{metstat}.

\section{OBSERVATIONS}
Our observations were made with  the Bell Laboratories 7 m telescope
in the spring of 1995. The beamwidth at 86 GHz is $2'$. System temperatures ranged from 300 to 400 K, but could be as high as 1200 K. Generally while
mapping, the observing time per point was adjusted to retrieve uniform rms
noise values. We observed the lines listed in
Table \ref{lines}, some of them (as listed in the table caption)
with a $256\times 1$ MHz filterbank, others with a
$256\times 250$ kHz filterbank.

\begin{figure}
\includegraphics[width=16cm]{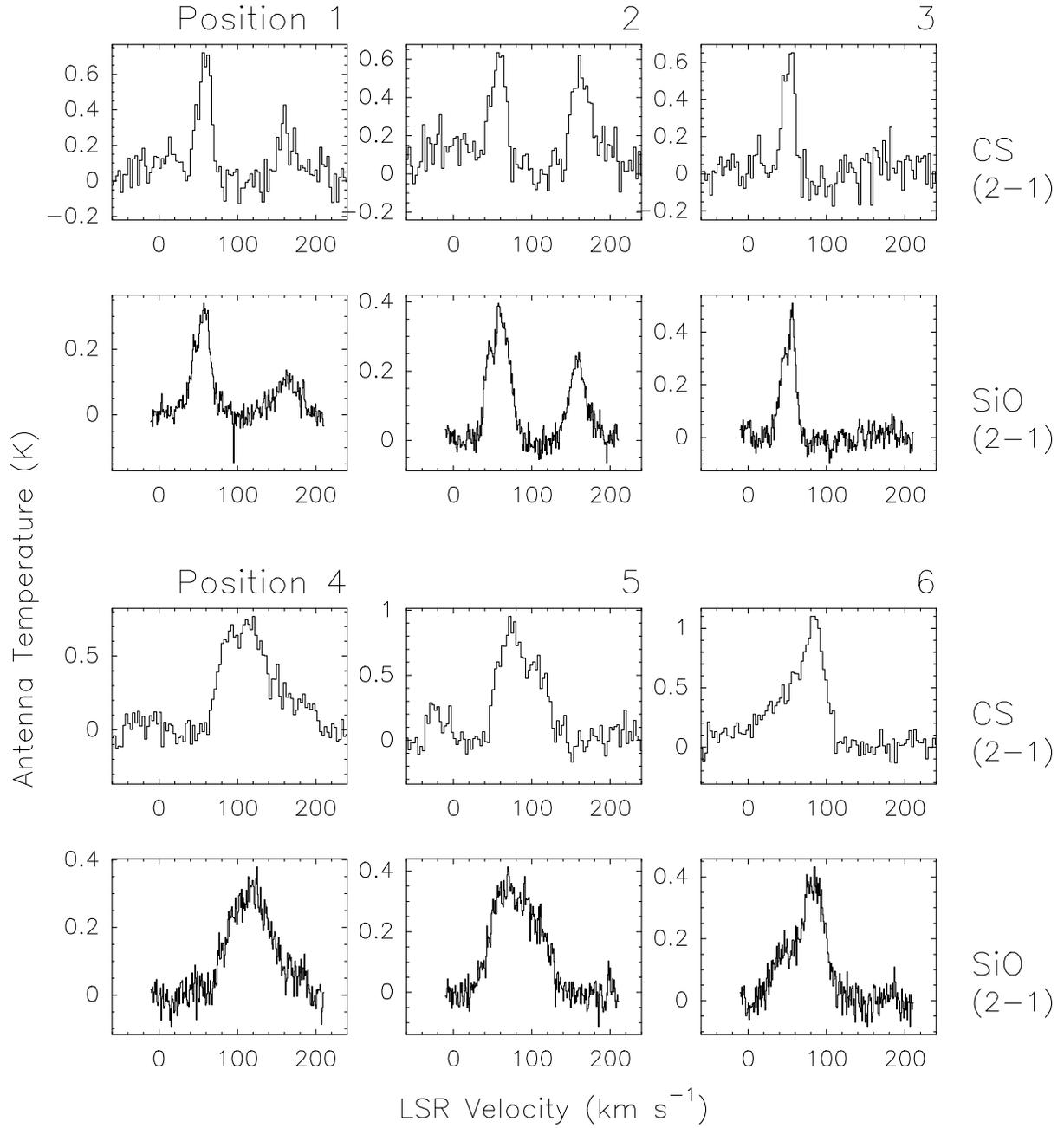}
\caption{
{\it Top to bottom}: Spectra taken toward positions 1--6 of
Table \ref{positions} in the $J=2-1$ lines of CS and SiO.}
\label{fig:siocsfig}
\end{figure}

\begin{figure}
\includegraphics[width=16cm]{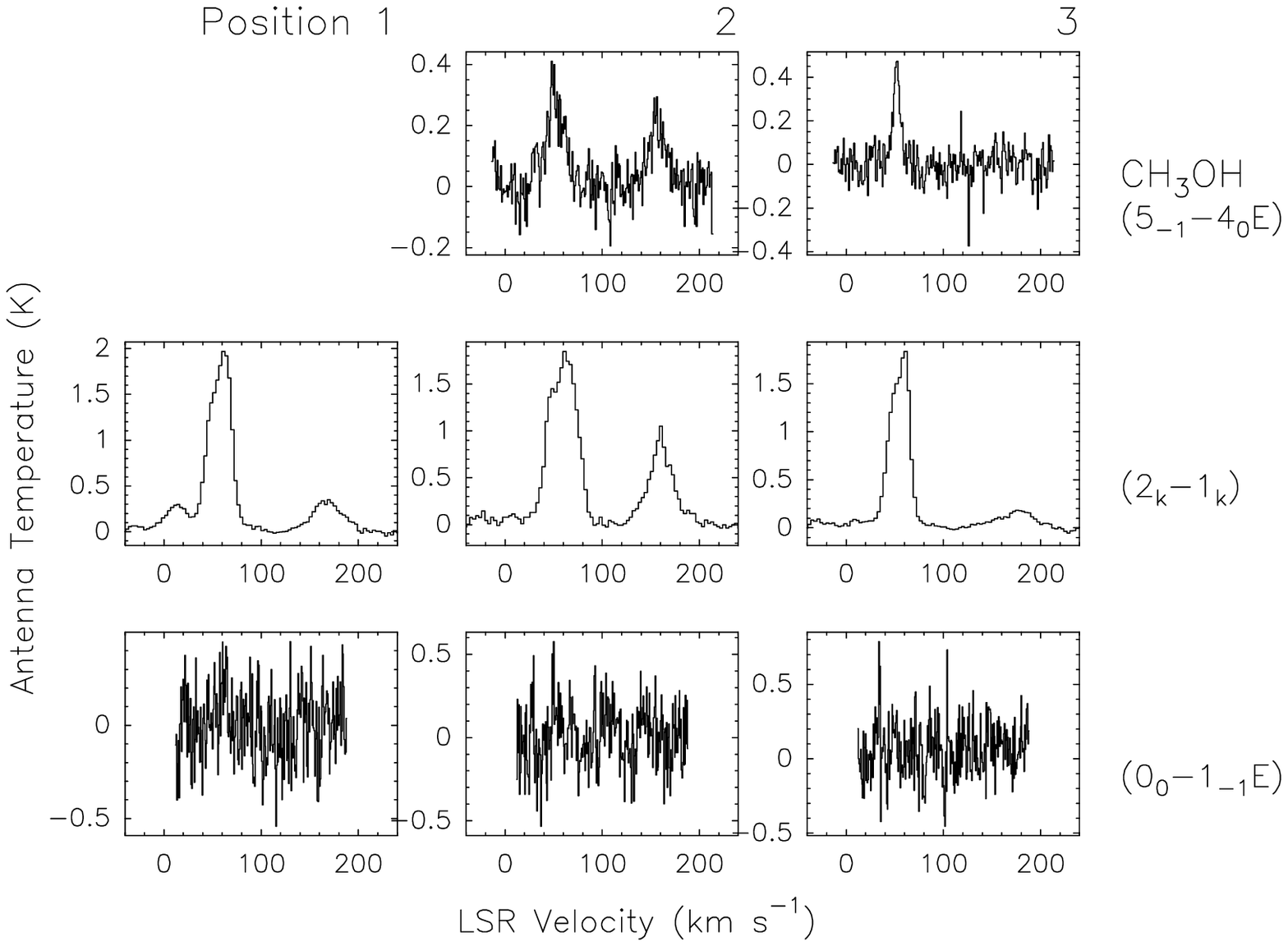}
\caption{
{\it Top to bottom}: \meth\ spectra taken toward positions 1--3
of the 84 GHz \EFGH\ line, the
$2_k-1_k$ quartet  near 97 GHz, and the 109 GHz
$0_0-1_{-1}E$ line. The 97 GHz quartet is a
blend of  three $E$-type  and one $A^+$ line (see Table \ref{lines})
and
the  LSR velocity scale is relative to the latter.}
\label{fig:ch3ohfig1}
\end{figure}

\begin{figure}
\centering
\includegraphics[width=16cm]{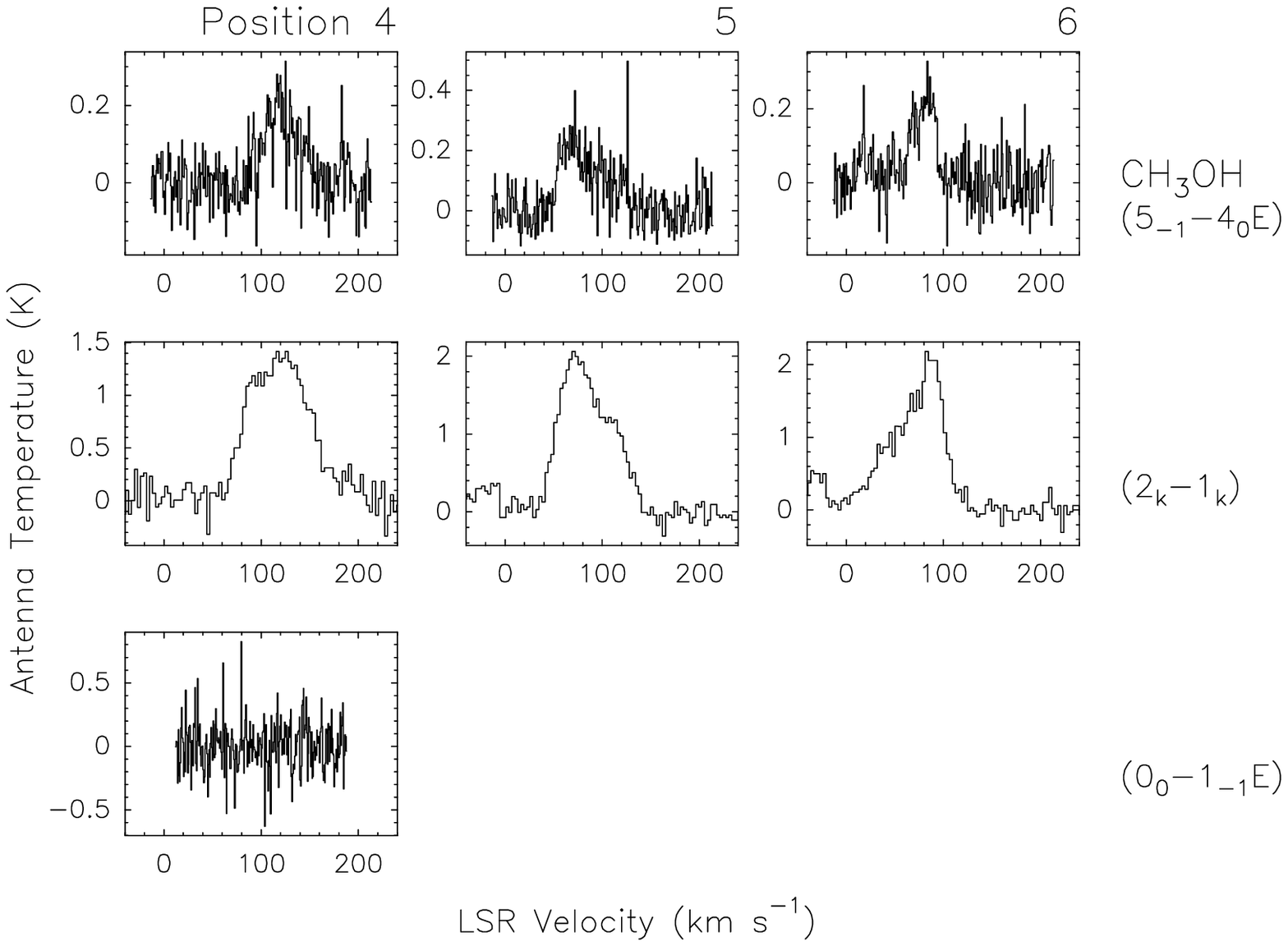}
\caption{
Same as Fig. \ref{fig:ch3ohfig1} for positions 4--6.}
\label{fig:ch3ohfig2}
\end{figure}

\begin{figure}
\includegraphics[width=16cm]{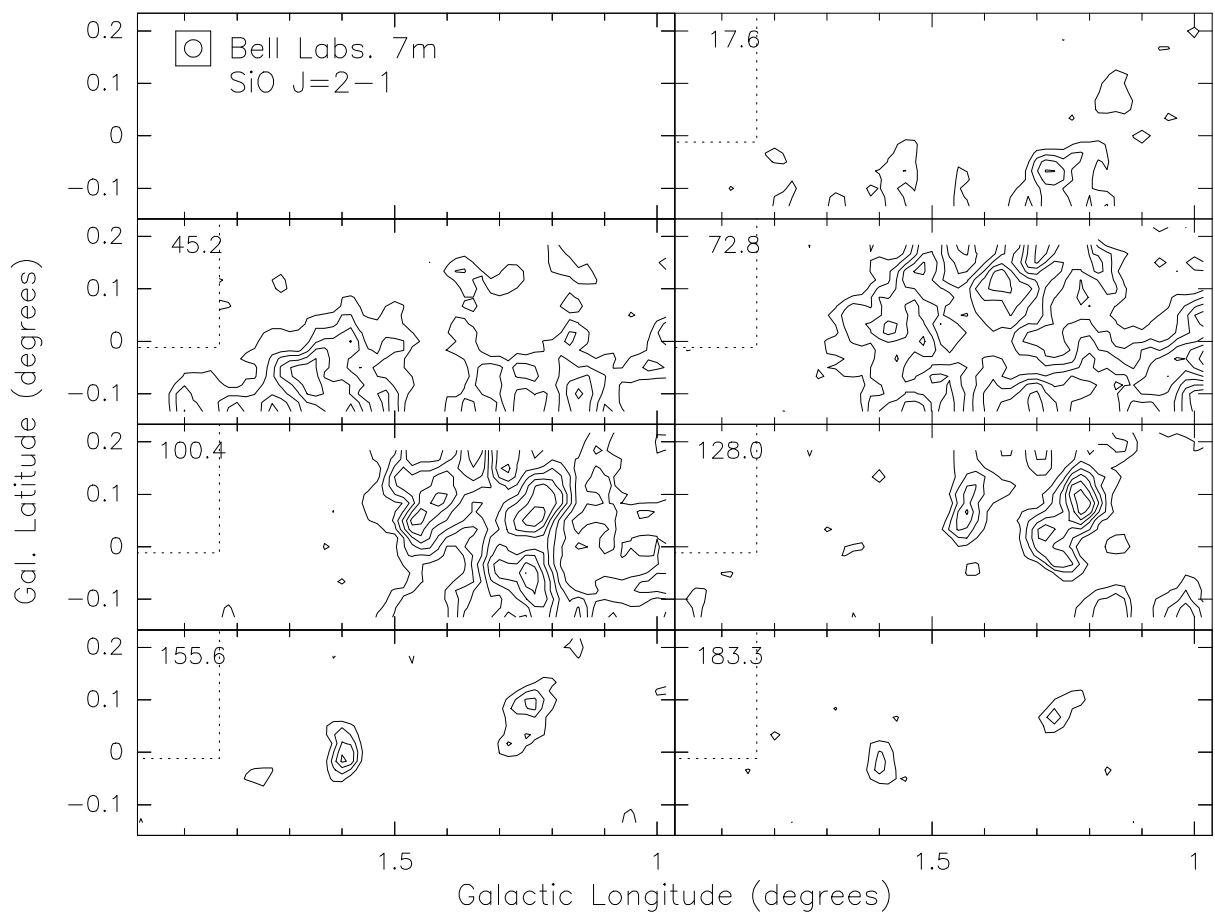}
\caption{
Maps of SiO $J=2-1$ emission of G1.6$-$0.025 made with the Bell Labs.
7m telescope. Each map represents the emission in \TAS\ units
in a  ''channel'' smoothed to a width of 27.6 \kms\ centered on the velocity (in \kms)
given in the left upper corner of each panel. Contours
are 2 to 20 in steps of 2 times 25 mK, which is equal to the
rms noise. The area within the
\textit{dotted lines} was not mapped. An additional area of width
$-0\decdeg11 < b <0\decdeg0$ was also mapped
from $l = 2\decdeg$ to $2\decdeg6$ but no emission was found within it.
The $2'$ diameter beam (FWHM)  is indicated in the \textit{left upper
corner} of the \textit{left upper panel}.}\label{fig:channelmaps}
\end{figure}

We mapped only  the SiO $(2-1)$ line extensively over
the $l,b \sim 1\decdeg0 \times 0\decdeg35$-sized region
shown in Figs.  \ref{fig:integratedmap} and \ref{fig:channelmaps}.
Mostly high-quality spectra of this line and the others
listed in Table \ref{lines} were taken toward the
``fiducial'' positions given in Table \ref{positions},
which are marked in Fig.
\ref{fig:integratedmap}.
The CS and SiO spectra taken toward positions 1--6
are presented in Fig. \ref{fig:siocsfig} and the \meth\
spectra in Fig. \ref{fig:ch3ohfig1} and \ref{fig:ch3ohfig2}.
Line parameters , obtained by Gaussian fitting
are presented in Table \ref{gauss}.
Because of the existence of multiple velocity components,
large line widths and line blending, Gaussian fitting of the \meth\
\zkok\ series was not viable. The interpretation of the
methanol results is discussed in \S \ref{metstat}.

\section{OUR MOLECULAR LINE DATA}
As shown in  Figs. 2--5, we observe emission between velocities
of $\sim -10$ and $+200$ \kms.
Molecular gas at velocities  $< -100$ \kms\ and $>120$ \kms\
is usually attributed to the expanding molecular ring \citep[EMR, ][]{Kaifu_etal1972,Kaifu_etal1974,Scoville1972} around the Galactic center.
Given its position, G1.6$-$0.025 is at the high longitude end of that
ring. That the emission in  all the three molecules
observed by us is much more prominent and widespread than  in ``normal''
Galactic disk clouds favors a
Galactic center location; see also the arguments brought forward in \S \ref{ewmg}.

\subsection{SiO Emission Distribution and Velocity Structure\label{MED}}
Figs. \ref{fig:integratedmap} and \ref{fig:channelmaps} show that the
SiO $J = 2 -1$ emission is very clumpy on different scales, the
smallest of which seem to be resolved by our beam, whose HPBW
corresponds to 4.7 pc. This picture is similar to that presented by
the $J = 1 - 0$ line mapped by \citet{Martin-Pintado_etal1997} with
the same resolution over a similar-sized region extending the area
mapped by us to smaller longitudes, i.e., from $l=+0\decdeg8$ to
$l=-0\decdeg2$, covering Sgr B2 and the Galactic center proper (Sgr
A). \citet{Huettemeister_etal1998} observed the $^{28}$SiO and the
$^{29}$SiO $J = 2 - 1$ and the $^{28}$SiO $5 -4 $ transitions toward
CS peaks found by \citet{Bally_etal1987}. They used LVG calculations
to model these two-line, two-isotopomer data to constrain density,
temperature and SiO abundance toward all these positions. Their data
are consistent with a hot $T>100$ K, low density medium $\sim10^{4}$
\ccm, in which, particularly, the higher-$J$ SiO transitions are
highly subthermally excited. One of the positions they observed, at
$l,b = +1.31$, $-0.31$, is within $\approx 2'$ of our position
6. Toward this position, they find the highest fractional SiO
abundance ($10^{-8}$) of all the 33 positions they observed, which are
spread over the whole of the CMZ. \citet{Huettemeister_etal1998}
invoke a shock origin for the elevated SiO abundance there and also
toward other locations for which models of the Galactic bar
gravitational potential predict cloud-cloud collisions (see \S
\ref{ccc} and \ref{shockchemistry}).

\subsection{Methanol Emission}
Toward all of the positions
listed in Table \ref{lines} the emission from the
\zkok\footnote{The projection of the angular momentum quantum number, $k$,
runs from $-J$ to $+J$ for $E$-type CH$_3$OH. For $A$-type CH$_3$OH a capital $K$ is used, with
$0 \le K \le J$.
When referring in one expression to levels from both species, lowercase $k$ is used.} quartet of \meth\ is stronger than that of the SiO line, and, amazingly, even stronger than that of the CS line
(see Table \ref{gauss}, Figs. \ref{fig:ch3ohfig1} and \ref{fig:ch3ohfig2}).
Given their strong blending, it is impossible to determine
the properties of these lines by fitting Gaussians in any
meaningful way.
Instead we performed the model calculations described in \S
\ref{metstat} to predict intensities for them
lines and the 84 GHz \EFGH\
transition and, in turn, to constrain the physical parameters of the emission region.

\subsection{\label{ccc}Evidence for Cloud-Cloud Collisions}
We note that we find a spatial coincidence
(at $l,b= 1.23,-0.05$) of an SiO clump with emission in the 17.6 \kms\ channel with one with emission in
the 100 \kms\ channel. We also observe a coincidence,
at $l,b=1.22,+0.10$, between clumps in the
100.4 and one in 155.6 \kms\ channels. Other such coincidences
can be found in the channel maps.

A similar coincidence of two clumps with widely
different velocities that appear at the same projected area in space
has been reported by  \citet{HaschickBaan1993} who note,
in the \TSGH\ \meth\ emission,
a coincidence of a clump in the 40--70 \kms\ velocity
interval (which they call Dm)
with one in the 150--167 \kms\ interval (Em) at $l,b=
1\decdeg594, +0\decdeg015$. This particular spatial coincidence of components at
these velocities was also pointed out by \citet{Salii_etal2002}, who present a map of the
high velocity emission along with the integrated emission in the  \EFGH\
and the blended $2_k - 1_k$
lines of \meth\ for which we took spectra only toward selected positions.
Both the low and the high velocity emission have a similar extent in the \meth\
as the SiO emission (for the high velocity emission, see the 156 \kms\ channel map in our Fig.\ref{fig:channelmaps}).

\citet{HaschickBaan1993} credit \citet{Sobolev1992}, who interpreted
the \TSGH\ \meth\ data of \citet{Berulis_etal1992} in a scenario
involving a cloud-cloud collision,
an intriguing idea, which is further promoted by \citet{Salii_etal2002}; see \S\ref{shockchemistry}.



\section{PHYSICAL CONDITIONS AND CHEMISTRY IN \go}
\subsection{Methanol Statistical Equilibrium Calculations}\label{metstat}
In addition to our own  data and those of \citet{Salii_etal2002}, further
important constraints on \meth\ excitation in G1.6$-$0.025 come from
widespread enhanced absorption (overcooling) in the 12.2 GHz
\etrans\ line \citep{WhiteoakPeng1989}
and the also widespread maser emisson
in the \TSGH\ line \citep{HaschickBaan1993,LiechtiWilson1996}. As explained, e.g., by
\citet{Menten1991} one expects, in the absence of
a strong far-infrared field, which is certainly the case
in G1.6$-$0.025, an overpopulation of the $k=-1$
ladder relative to the $k=0$. Similarly, one
might naively expect overcooling in the 109 GHz  $0_0-1_{-1}E$
transition, where we observe neither absorption
nor emission with a $3\sigma$ upper limit of 0.57 K. This absence
is explained by our model predictions (see below).

To address these issues quantitatively, we performed model calculations.
\citet{Leurini_etal2004} used the rate
coefficients for collisions of CH$_3$OH with He calculated by
\citet{Pottage_etal2002} to investigate
the excitation of CH$_3$OH over a range of physical
parameters typical of star-forming regions. They also
presented a new technique
to handle the problem of deriving physical parameters of
a source from spectroscopic data; the technique is based on
the simultaneous fit of multiple lines in a spectrum (when present) with a
synthetic spectrum computed using the
LVG approach for solving the
radiative transfer equations, in the derivation of
\citet{deJong_etal1975}. This analysis is particularly
well-suited for the case of strongly blended lines, where
``by-hand'' Gaussian fitting
of lines with multiple components often leads to unreliable results.
Moreoever, following \citet{CesaroniWalmsley1991},
the effect of overlap of lines in the excitation of the
CH$_3$OH molecule
is taken into account defining an average
optical depth and brightness temperature for lines with  a frequency
separation
\begin{equation}\label{lineoverlap}
\nu_{i}- \nu_{j} \le \Delta\nu_{i}+\Delta\nu_{j}
\end{equation}
Using the technique described above, we analyzed our data toward
two of the observed positions, namely numbers 2 and 3 of Table~\ref{positions}, to derive
the physical parameters of the region. In addition to our own data, we have also compared
the predictions of our model for other  CH$_3$OH lines observed toward position 2 by other authors.
Our model calculations do not predict absorption in the $0_0 - 1_{-1} E$
transition for a wide range of physical parameters $(n $ from $10^3$
to $10^8~\rm{cm}^{-3}$; $T_{kin} $ from $ 10 $ to $ 200~\rm{K}$;
$N(\rm{CH_3OH})$ from $10^{12} $ to $ 10^{16}~\rm{cm}^{-2})$.
The simultaneous modeling of the $2_k - 1_k$ and $5_{-1} - 4_{0}E$ lines gives constraints on the
column density of methanol and on the H$_2$ density. The $5_{-1} - 4_{0}E$ transition is strongly inverted
 over a wide range of physical parameters, for $n($H$_2) > 10^4$ \ccm\ and
CH$_3$OH column densities higher than 10$^{15}$ cm$^{-2}$ (see Fig.~\ref{tau84}). Hence, the non-detection of
obvious strong maser action (but see below) in our observations indicates low values for the column density of the gas.
The $2_k - 1_k$ lines are, on the other hand, sensitive to the density of the gas (see Fig.~4 of \citet{Leurini_etal2004}).

\begin{figure}[htb]
  \centering
    \includegraphics[angle=-90,width=9cm]{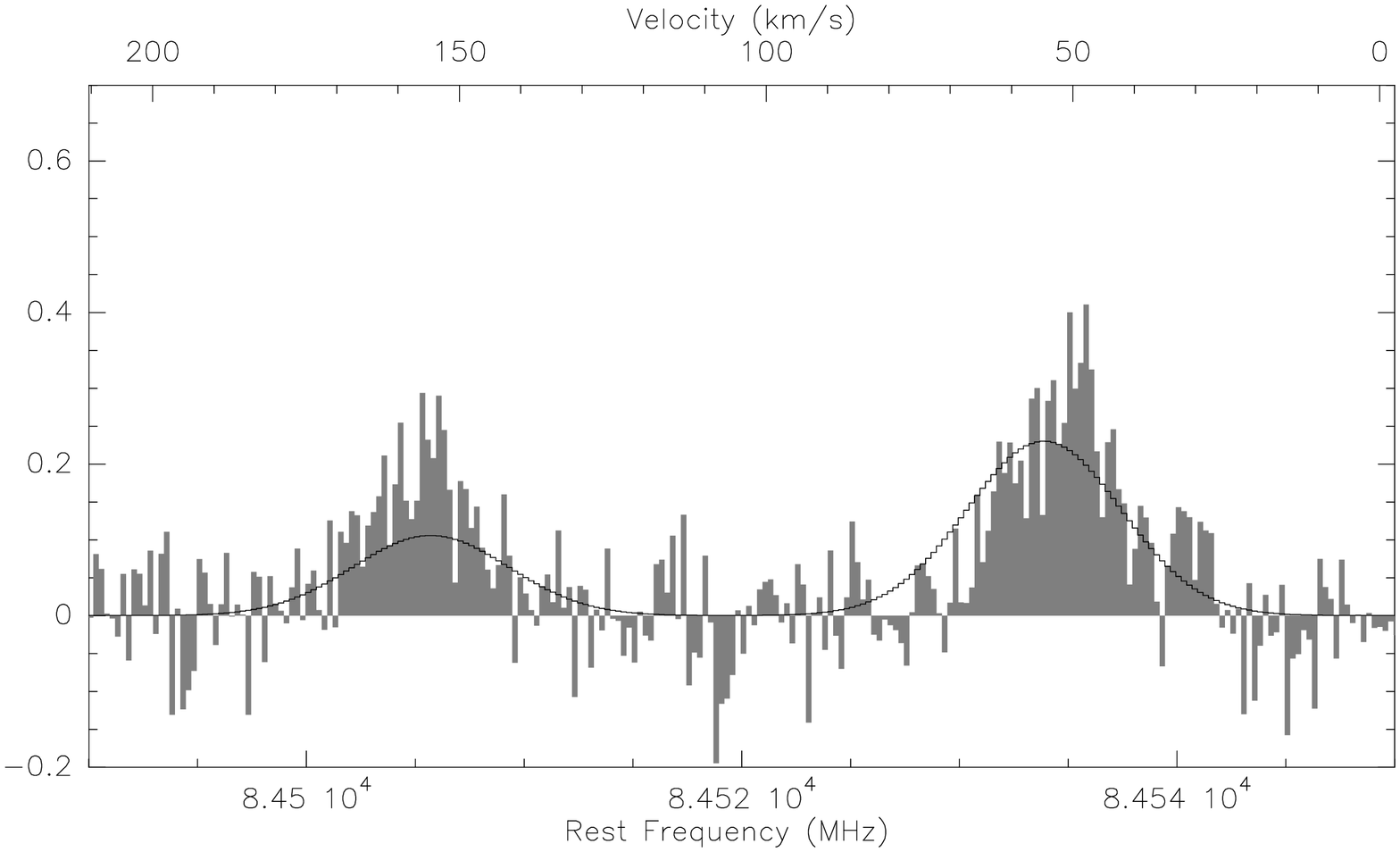}
    \includegraphics[angle=-90,width=9cm]{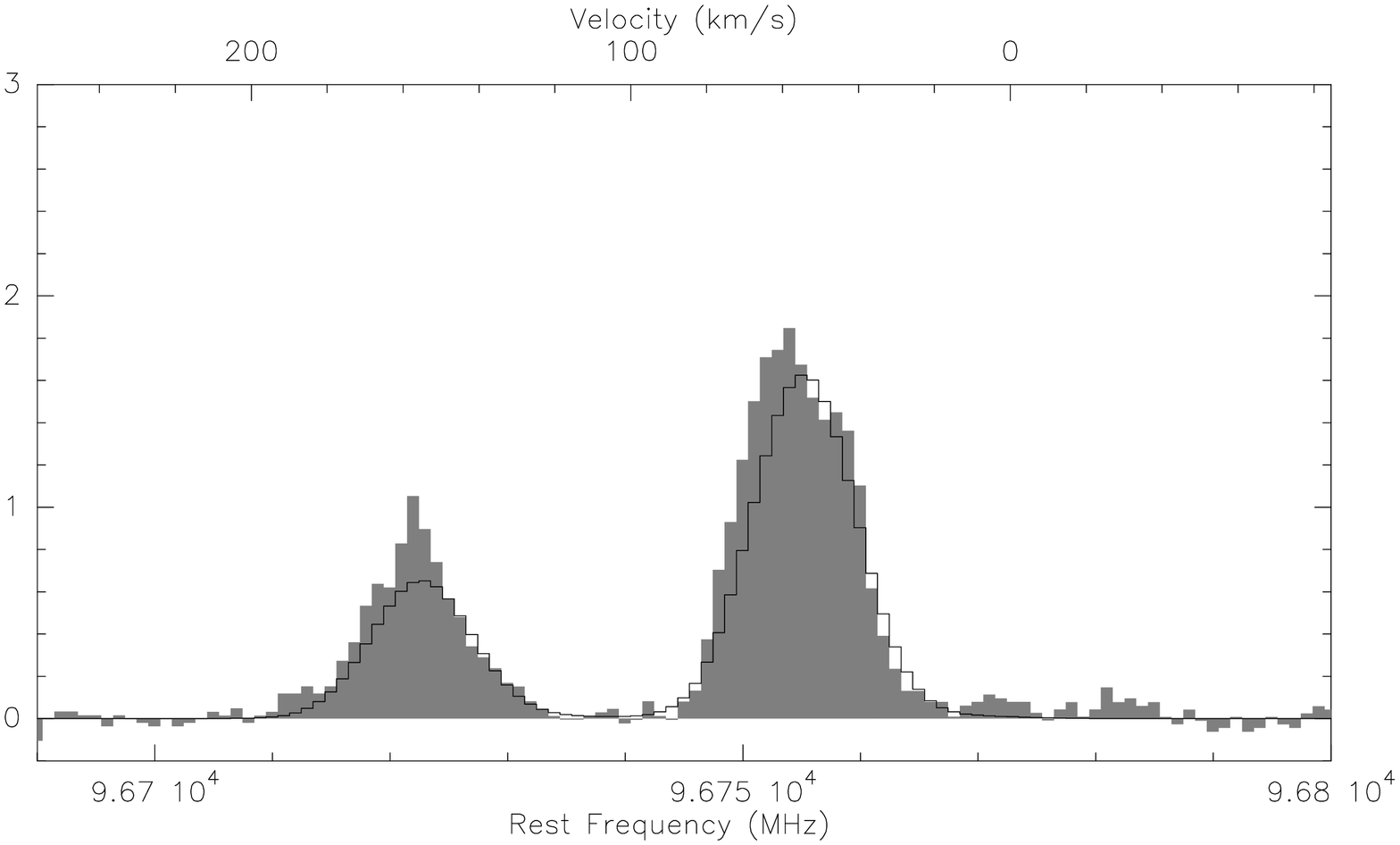}
\caption{Best LVG  fit  toward position 2 overlaid on the real data for the 5$_{-1} - 4_{0} E$ line \textit{upper panel}
and the $2_k - 1_k$ quartet of lines \textit{lower panel}.}\label{fit3}
\end{figure}

\begin{figure}[htb]
    \includegraphics[angle=-90,width=9cm]{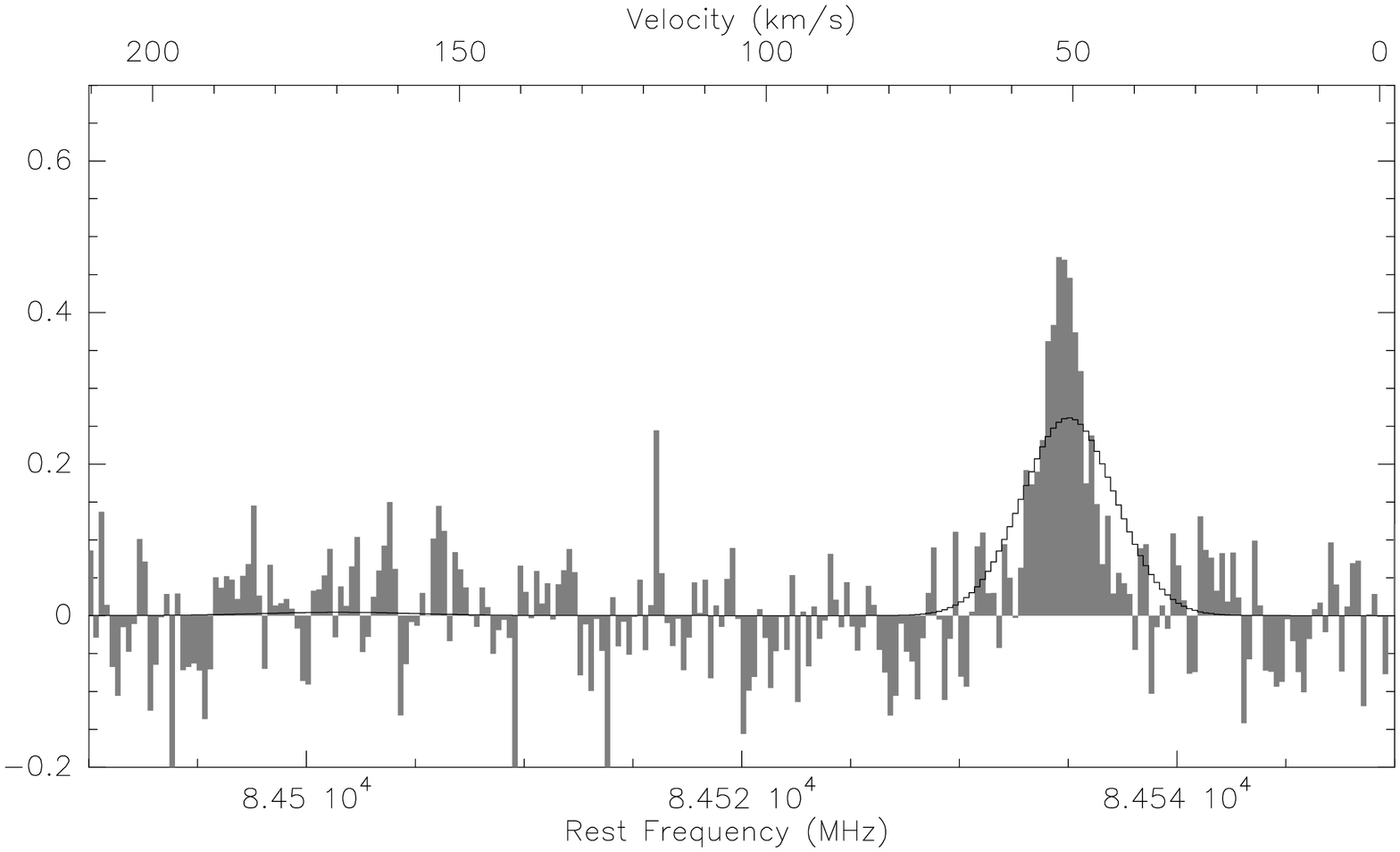}
    \includegraphics[angle=-90,width=9cm]{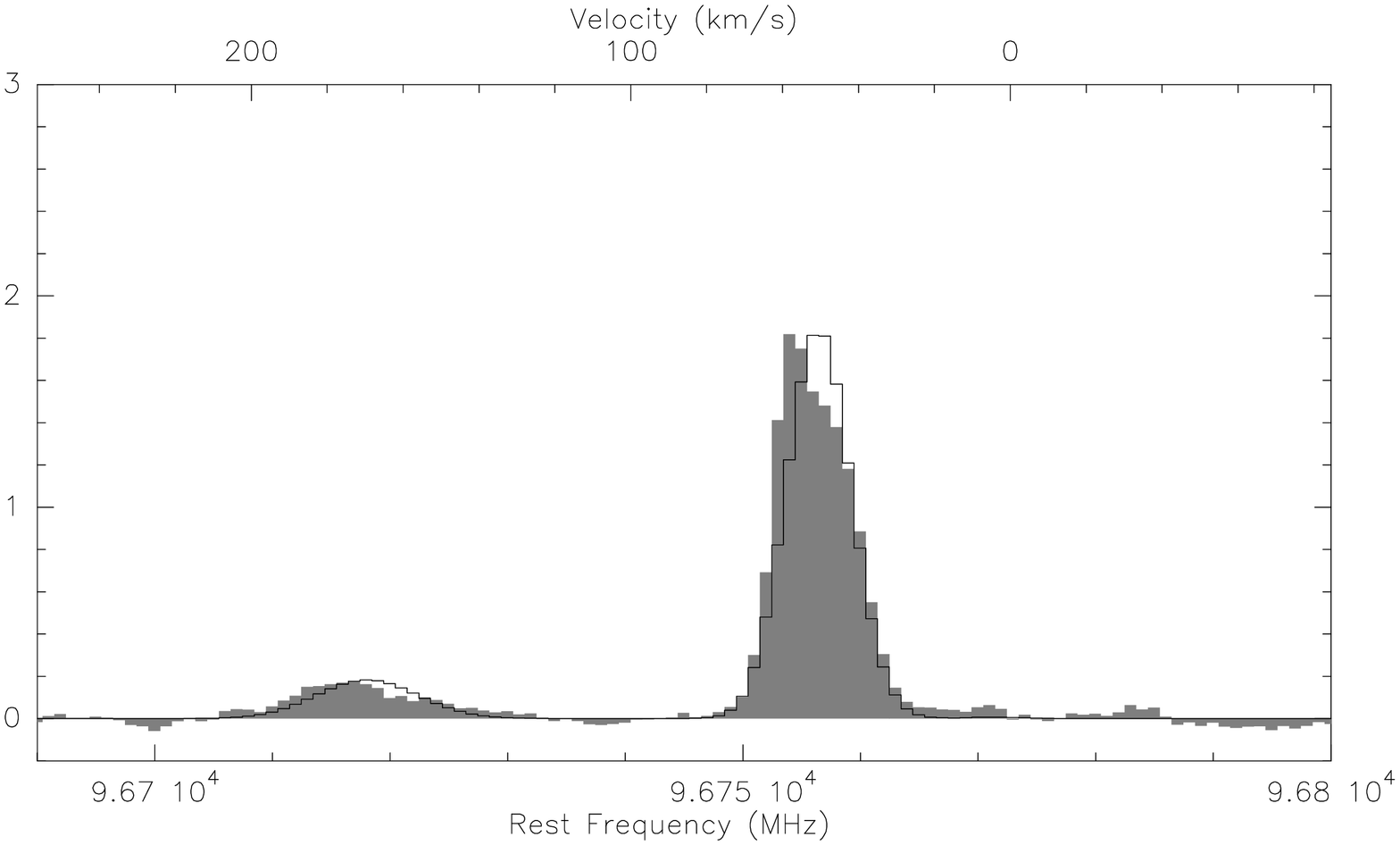}

\caption{Same as Fig.~\ref{fit3} for position 3.}\label{fit4}
\end{figure}

Inspecting the $2_k - 1_k$ CH$_3$OH map presented by \citet{Salii_etal2002} we assumed
the CH$_3$OH emission to be extended compared to the 7~m telescope beam. We modeled the data with two velocity
components, corresponding to the high velocity clump and to the extended cloud.
The LVG fit overlaid on the data is shown in
Figs.~\ref{fit3} and \ref{fit4}.
The results determined from the
fit are given in Table~\ref{para}.
Table~\ref{modeltable} shows our model
predictions for the other CH$_3$OH transitions observed by other authors.
Pointing and absolute calibration uncertainties and beam-size differences
can make the determination of
physical parameters less reliable when comparing data from different
telescopes; in our case however,
beam-size differences should not affect the results as, given the
source sizes, beam filling factors for the
different transitions are pretty close to 1.

\citet{Salii_etal2002} determined the physical parameters
toward our position 2 by analyzing several methanol transitions. Their
results do
differ
somewhat from ours.
In particular, they find
somewhat smaller spatial densities and \meth\ column densities. From their
analysis they conclude that the high velocity clump has a hydrogen density, $n($H$_2)$,
less than $10^4$~cm$^{-3}$, column densities between $4\times 10^{11}$
and $6\times 10^{12}$~cm$^{-2}$ and temperatures in the 150--200~K
range.  For the extended cloud they infer a hydrogen density
$<10^6$~cm$^{-3}$, column densities larger than $6\times
10^{11}$~cm$^{-2}$ and a kinetic temperature of less than 80~K.  Assuming
that all the gas is in molecular form, this translates in spatial
densities of molecular hydrogen less than $3\times 10^3$~cm$^{-3}$ for
the extended cloud and $n($H$_2)<5\times 10^3$~cm$^{-3}$ for the high
velocity clump. With these parameters, they can reproduce the observed
line intensities for several transitions, but fail to explain the simultaneous
deep absorptions in the 2$_0 - 3_{-1} E$ and 2$_1 - 3_{0} A^+$ lines  at 12.18~GHz and 156.6~GHz,
and the brightness of the emission in the $J_0 - J_{-1} E, J=1,2,3$ blend.
They conclude that the absorptions and
the $J_0 - J_{-1} E, J=1,2,3$ blended emission come from different parts of the
cloud.

Our model predictions (see Table~\ref{modeltable}) overestimate the
 $J_0 - J_{-1} E$, $J=1,2,3$ blended emission and
underestimate the absorption in the 2$_0 - 3_{-1}$-$E$ transition in
the extended cloud, but can reproduce the simultaneous absorptions and the emission in the $J_0 - J_{-1} E$ blend.
For the high velocity clump, our predictions are in good agreement with the observations.
Moreover, spatial densities of
the order of a few 10$^3$~cm$^{-3}$, as inferred by \citet{Salii_etal2002}, fail  to reproduce our
observations of the $5_{-1} - 4_{0}$-$E$ line.

At position 2, \citet{HaschickBaan1993} find maser emission in the
4$_{-1} - 3_{0}$-$E$ line on top of a broad thermal-looking component. Fitting
the maser component in our model would need a second, narrower
component for both the high velocity clump and the extended cloud,
which is not detected in our observations. Therefore we do not include
in our analysis any other component to account for the narrow
maser features in the 4$_{-1} - 3_{0} E$ line.  However, the maser
action in the 4$_{-1} - 3_{0} E$ and the simultaneous absence of it in
the 5$_{-1} - 4_{0} E$ line can give interesting constraints for the
physical parameters of the regions.  Leurini et al. (in prep.) have
extensively analyzed the pumping mechanisms of class I CH$_3$OH
masers.  Their calculations confirm collisions to be responsible for
the excitation of class I masers and suggest the maser action in these
lines to be used as a density indicator.  Both the 4$_{-1} -
3_{0}$-$E$ and 5$_{-1} - 4_{0} E$ transitions are inverted at low
densities; however, the inversion of 5$_{-1} - 4_{0} E$ line starts
with $n($H$_2) > 10^4$ \ccm\ with CH$_3$OH column densities higher
than 10$^{15}$ cm$^{-2}$, while the 4$_{-1} - 3_{0} E$ line mases also
at lower densities, as Fig.~\ref{tau} shows, almost independently
from the kinetic temperature. The above is also found by
\citet{Berulis_etal1992} and \citet{Sobolev1992}. At lower column
densities, the inversion of the 5$_{-1} - 4_{0} E$ line happens at
slightly higher densities.  Therefore the detection of a maser
component at 36~GHz together with the non-detection of strong
maser action manifested by narrow features in the 84.5~GHz line puts
an upper limit on of $\sim 10^4 - 10^5$ cm$^{-3}$ on the spatial
hydrogen density of the region, depending on the column density of
methanol.

Nevertheless, comparison of 5$_{-1} - 4_{0} E$ and $2_k
- 1_k$ spectra (Figs. \ref{fig:ch3ohfig1} and \ref{fig:ch3ohfig2})
shows that the former generally are narrower than the latter, which
might at first look be interpreted as line narrowing accompanying
maser action. However, since some of the (noisy) spectra also cover
incongruent velocity ranges this might actually not be the cause for this
difference in appearance. Another possibility would be the existence
of two gas components in the beam with different densities and/or
temperatures, which might contribute to the lines in question in
different proportions.
This could explain the narrower line profiles of the 5$_{-1} - 4_{0} E$ line, which seems not to be excited in one of the velocity components, and the underestimate of line intensity in the model. However, the maps
of the 5$_{-1} - 4_{0} E$ and $2_k-1_k$ lines published by \citet{Salii_etal2002} (and with a spatial resolution of $\sim57''$), do not show any large discrepancies between the distributions of the two lines.
Needless to say, the described discrepancy illustrates the qualitative nature of our results.

To summarize, our statistical equilibrium modeling of the \meth\ lines
observed by us and others indicates relatively high densities ($>6
\times 10^4$~cm$^{-3}$) and moderate temperatures ($30$ -- $60$~K) for
two representative positions in the G1.6$-$0.025 cloud at
v$_{\rm LSR}$=50~km~s$^{-1}$. In the high velocity component, lower densities
are inferred by the model for both positions; for position~2, a high
temperature (190~K) is needed to reproduce the observations, while
cold gas (16~K) is needed for position 3. The latter is puzzling as,
in analogy to position 2, one might also expect an enhanced
temperature for high velocity \meth\ emission. 
High velocity SiO is not even detected toward
position 3 and high velocity CS barely.  Methanol abundances relative
to H$_2$ are of the order of 10$^{-7}$ -- 10$^{-8}$. Finally, we note
that the temperatures and densities we derive are of the same order of
the values that \citet{WhiteoakPeng1989} derive from their modeling of
the \meth\ \etrans\ line.

\citet{Huettemeister_etal1993} conducted a multi-transition study of
\nhhh\ inversion lines with energies of up to 408 K above ground
toward 36 positions in the CMZ. They find evidence for two components
pervasive throughout the CMZ, both at each position at roughly the same
velocity, a cool one, $T \sim 25$ K and a hot one $\sim 200$ K. The
densities of the these components are not well constrained by their
observations, but they argue that the hot and cool gas have densities
of $10^4$ and $10^5$ \ccm, respectively.

In their study, they also included one position within the area that
we mapped in SiO emission ($l,b = +1\decdeg15, -0\decdeg09$) and
several others just abutting it; i.e., within 0.2 degrees outside of
its boundaries. Since they do not present data toward the positions
toward which we see a high(=clump)- and a low(= general
cloud)-velocity component, a direct comparison with our analysis is
difficult. However, we can say that we do not see evidence for a high
temperature component at the general cloud velocities toward the two
positions that we analyzed, while they find evidence for both a cool
and hot component at these velocities.

The reason why we do not see a hot component may be a selection effect
based on our choice of lines. The rotation temperatures
\citet{Huettemeister_etal1993} derive from their lowest excitation
lines alone [$J,K$ $=(1,1)$ and $(2,2)$ at energies of 23 and 64 K above ground, respectively]
invariably are between 18 and 40 K throughout the CMZ and mostly
around 25 K. In contrast, rotation temperatures determined from the
(4,4) and (5,5) lines (at 200 and 295 K above ground, respectively) are all $>50$
K, most $>80$ K, and some as high as 200 K. Since we didn't observe
any high excitation lines, given the experience with \nhhh, we could
not detect any hot component in \go\ at velocities at which emission
from cold as well as from hot gas is arising, but only at velocities
at which only the hot component emits (i.e., high velocity clumps).

To estimate the column density of molecular hydrogen and derive the
[CH$_3$OH]/[H$_2$] ratio, we used the $^{13}$CO $J = 1 - 0$ data
imaged by \citet{Bally_etal1987} with the Bell Labs 7~m
telescope. Assuming the Local Thermal Equilibrium approximation, and
using the kinetic temperatures derived from the analysis of the
methanol emission, we computed the $^{13}$CO column densities for
positions 1 and 2, for the high velocity clump and for the extended
cloud, by integrating over the velocity channels corresponding to the
CH$_3$OH emission. To convert the $^{13}$CO column density into a
H$_2$ column density, we used a typical abundance relative to H$_2$ of
10$^{-4}$ for CO and a value of 20 for the $^{12}$C$/^{13}$C isotopic
ratio.
This is the value \citet{WilsonRood1994} give for the
$^{12}$C/$^{13}$C ratio for molecular clouds in the Galactic center
region. If we, however, used 11 for the latter ratio, which
\citet{GardnerWhiteoak1981} derive for G1.6$-$0.025 from observations
of formaldehyde isotopomers, the relative methanol abundance ratios
given in the following would increase by a factor of 2.

Assuming that the column density of CH$_3$OH-$A$ and CH$_3$OH-$E$ are the same,
our derived column densities translate into abundances relative to
molecular hydrogen of $10^{-8}$ for the high velocity clump and
$1\times 10^{-6}$ for the extended cloud toward position 2 and to $10^{-8}$ and
$3\times 10^{-8}$, respectively, for position 3.  The high velocity
component is not detected in the $^{13}$CO line toward position 3 and
we could only derive an upper limit to the H$_2$ column density based
on the rms noise value of the data. Therefore, the CH$_3$OH abundance ratio
estimated for this component is a lower limit to its true value.

\subsection{\label{shockchemistry}Shock Chemistry in \go}
As discussed in \S\S \ref{MED} and \ref{metstat} the SiO and \meth\ abundances are enhanced relative to cold molecular cloud values.
The observed chemical peculiarities of \go\ could be the the result of several
mechanisms all involving shocks: cloud-cloud collisions (see \S \ref{MED}), a molecular cloud-SNR
interaction, and whatever is responsible for the peculiar
large linewidths in GC GMCs in general (increased turbulence, tidal forces).

It is tempting to assume that the high observed SiO and \meth\
abundances in \go\ have a common origin. With regard to these two
species, a picture similar to \go\ presents itself in the molecular
peak M$-0.02-0.07$ (the so-called ``20 \kms\ cloud''), which lies
$\sim 2'$ NEE of the Galactic center radio source Sgr A$^\star$ at a
position where the supernova remnant Sgr A East (SNR G0.0+0.0) appears
to interact with a molecular cloud.  Here,
\citet{Martin-Pintado_etal1997} find strong SiO ($J=1-0$) emission,
while \citet{Szczepanski_etal1989} and, at higher resolution,
\citet{LiechtiWilson1996} find very strong maser emission in the 36
GHz \TSGH\ line of methanol. Just like in \go\ the \meth\ emission
consists of a few narrow spikes and intense ``broad'' emission. 1720
MHz OH masers are also found in this region on the near-side of the
SiO/\meth\ emission distribution relative to Sgr A East facing the SNR
\citep{Karlsson_etal2003,PihlstromSjouwerman2006,Yusef-Zadeh_etal2007}.
We emphasize the similarity to \go, with the SNR $1.4-0.1$ (see
\S\ref{SNR}), projected on and possibly interacting with it.

The existence of 1720 MHz OH maser emission argues, as discussed in \S
\ref{SNR}, also for (C-)shock.  The densities and temperature we
derive from the methanol lines are very similar to the values
necessary for 1720 MHz OH maser emission as discussed there.  The
chemistry in the molecular gas interacting with the SNR IC 443 was
studied by \citet{vanDishoeck_etal1993}. One subregion, clump G I, was
found to show a particularly rich chemical picture. However, one
dramatic difference between the molecular content of \go\ and
molecular IC 443 clump G I a is the complete dearth of any \meth\ in
the latter.  \citet{vanDishoeck_etal1993} give a relatively sensitive
$2\sigma$ upper limit of 0.2 K for any line in the 241.7 GHz \meth\
$5_k-4_k$ series, while various SiO lines are two to three times
stronger than that.

As for SiO, high abundances after the passage of magnetohydrodynamic
C-shocks can result from a combination of the setting free of SiO into
the gas-phase by sputtering of the (charged, and hence
coupled to the magnetic field) grain cores by neutral particles in the
region of the C-shock where the relative velocities between charged
and neutral particles are large, and following gas-phase reactions
\citep{Schilke_etal1997}.  This is true for pristine material, where silicon
resides in the grain cores.  If, as is the case in the Galactic
center, molecular clouds are frequently exposed to cloud-cloud
collisions, silicon may, after the initial release from the core,
reside either in the gas phase (possibly some of it in SiO$_2$ as
suggested by \citet{Schilke_etal1997})
or in grain mantles, with a lower binding energy.

How is the methanol produced in shocks?
Gas-phase production of methanol has been shown to be insufficient
to create the observed abundances in dark, quiescent clouds by many orders
of magnitude \citep{Geppert_etal2005}. 
Grain surface
production however seems to be efficient, as shown by
\citet{Hidaka_etal2004}.  This indeed is the proposed mechanism for
producing the high methanol abundances found in ``hot cores'' around
high-mass young (proto)stars, where \meth\ is created by the
evaporation from grain mantles once temperatures exceed $\sim 100$ K.
Shocks also would be able to release material from grain mantles, either by
sputtering, even at lower shock speeds than needed for SiO production,
since here the more weakly bound ice mantles have to be destroyed
instead of the grain cores, as needed for SiO; or by thermal
evaporation in the hot shocked gas.  This mechanism has indeed been
evoked for explaining the high methanol abundances in shocked protostellar outflows \citep[e.g.,\ ][]{BachillerPerez1997}.

To evaporate methanol from ice mantles, however, these ice mantles
have to be present.  While there is observational evidence from IR spectroscopy
that ice mantles do
exist in the cold envelopes of protostars (to be released by
outflows or heating by the igniting star), the conditions for formation of ice mantles in the highly
turbulent, warm and relatively low density environment of Galactic
center clouds (of which \go\ is a member) seem less favorable.
Deciding if sufficient methanol abundances in ice mantles can be built
up under these conditions will require detailed modeling.

Another option could be that under GC cloud  conditions, ice mantles do
not build up, but CO can reside long enough on the grain surface to be
transformed into \meth, which then is desorbed either by sputtering in
vortices, or just thermally desorbed. In this case, the elevated
methanol abundance would not be related to any specific shock event,
but be the steady state abundance under these special conditions.
Modeling would certainly be illuminating, but an observational
consequence of this mechanism would be a uniformly high methanol
abundance in the Galactic center, which may actually apply \citep{Requena-Torres_etal2006}.

Alternatively, in the elevated temperatures of a C-shock, \meth\ may be
created by the endothermic gas-phase reactions
\begin{equation}
{\rm CH}_4 + {\rm OH} \rightarrow {\rm CH}_3{\rm OH} + {\rm H}
\end{equation}
and
\begin{equation}
{\rm CH}_3+{\rm H}_2{\rm O} \rightarrow {\rm CH}_3{\rm OH} + {\rm H}
\end{equation}
discussed by \citet{Hartquist_etal1995}, which have
endothermicities of 6500 and 14700~K, respectively. 
The abundances of reaction partners OH and H$_2$O certainly would be
enhanced in a shock, but to judge these reactions' importance would require running
shock models, looking also carefully at possible shock destruction
mechanisms for \meth. Observationally, this would, just as the option
of removing \meth\ from ice mantles, imply a correlation of elevated
\meth\ abundances with shock events.

In practice, it will be very hard to distinguish observationally
between these scenarios, because SiO, the canonical shock tracer, is so
widely distributed.  This suggests that either shocks are ubiquitous,
or that the grains are processed to a degree that a significant
fraction of silicon resides in more weakly bound form either in the
gas phase or on the grain surface, so that the release mechanisms for
SiO and \meth\ are similar.

While we argue above that the interaction with a SNR may influence the chemistry in at least part of \go, cloud-cloud collisions resulting from the special dynamics induced by the Galactic bar potential may play a major role for \go\ and other GC clouds as a whole \citep[see][]{Huettemeister_etal1998, Rodriguez-Fernandez_etal2006}.

\section {\label{cms}G1.6$-$0.025 IS AT THE OUTER REACHES OF THE GALACTIC CENTER ORGANIC MOLECULE REGION}
There actually is evidence for a giant repository of organic molecules
coextensive with the Central Molecular Zone, of which G1.6$-$0.025 demarcates
the high-longitude border \citep{Menten2004}.

The first evidence for extended organic material in the CMZ
came from widespread 4.8 GHz H$_2$CO absorption
\citep{Scoville_etal1972}; see also \citet{Zylka_etal1992}. Given the ubiquity of formaldehyde
in molecular clouds \citep[e.g.][]{Downes_etal1980},
 one might dismiss this ``as nothing
special''. CH$_3$OH, on the other hand, has usually quite
low abundance and is difficult to detect outside hot, dense cloud cores.
Nevertheless, \citet{Gottlieb_etal1979} find the 834 MHz
$(1_1-1_1)A^\mp$ line in the CMZ in emission and \textit{extended}
relative to their $40'$(!)
beam,  concluding it is (weakly) inverted and amplifying the strong
background radio emission.

Other molecules similarly (or even) more complex than \meth\
and \hhco\ were found widespread throughout the CMZ, such as
HCOOH and C$_2$H$_5$OH
\citep{Minh_etal1992,Martin-Pintado_etal2001} and, very recently,
(CH$_3$)$_2$O, HCOOCH$_3$, HCOOH, and CH$_3$COOH \citep{Requena-Torres_etal2006}

Furthermore, mapping of the HNCO $5_{05}-4_{04}$ transition (made serendipitously
simultaneously with a C$^{18}$O survey), shows that the emission in this
line is extending continuously from $l = -0\decdeg2$ to $+1\decdeg7$, right
out to \go!
\citep{Dahmen_etal1997}.
The possible existence of such a huge mass of organic material
in our Galactic center is extremely exciting and its extent, chemistry,
and excitation should be studied with suitable tracers.

In fact, \go\  coincides with the third-strongest
peak in the integrated HNCO distribution (after the general Sgr B2
region and an area around $l = 1deg, b= 0\deg$) and the ratio of the
integrated intensities of the HNCO to that of the C$^{18}$O line in \go\
is the highest in the whole CMZ. Maybe it the low UV radiation field
density  resulting
from the absence of young  high mass stars in \go\ is conducive to the
existence of (fragile) complex molecules.  Given this, one might also
expect other complex molecules than \meth\ and HNCO
to have large abundances in \go. Possibly the best spectral range to
search for those is the 3 mm window since, given our density estimates
(\S \ref{metstat}), submillimeter lines might have prohibitively high
critical densities.

\section{CONCLUSIONS}
Here we summarize our main conclusions. We find very little evidence for
star-formation in over the whole $\sim 0.2$ deg$^2$ region of the \go\ GMC.
Our large scale, medium resolution ($2'$)
mapping in the $J = 2 -1$ transition of SiO reveals clumpy
emission over an $\sim 0\decdeg6 \times 0\decdeg3$ region stretching along the Galactic
plane. Toward selected positions, we have observed emission in the
$2_k-1_k$ quartet of \meth\ lines and the CS $2 - 1$ line.
Toward all of these, the \meth\ lines are several times stronger
than both the CS and the SiO lines. In addition, spectra of other methanol lines where taken.
The wide widths of all the observed lines, similar
to values generally found in the Galactic center,
indicates a high degree of turbulence. A high velocity clump with a
$\sim 100$ \kms\ higher velocity than the molecular cloud may indicate a
cloud-cloud collision.
Statistical equilibrium modeling of all the
\meth\ lines observed by us and others indicates relatively high
densities and moderate temperatures for one representative position in the
G1.6$-$0.025 cloud ($8~10^4$ \ccm/30 K) and higher temperature (190 K),
but lower density ($4~10^4$ \ccm in the high velocity clump.
For a second position 
we also find densities of several times $10^4$~cm$^{-3}$ for both the
low and the high velocity emission, but a puzzling low temperature
for the high velocity clump (16~K) and warmer gas (60~K) for the low velocity
50~$km~s^{-1}$ gas. Different
scenarios are possible in which shock chemistry might enhance the \meth\ and
SiO abundances in G1.6$-$0.025 and elsewhere in the Central Molecular Zone
by grain-gas chemistry or by hot gas chemistry.

\acknowledgments{We would like to thank Darek Lis for fitting the FIR/submm spectral
energy distributions, Maria Messineo and Harm Habing for
discussions of OH/IR stars, and Malcolm Walmsley for comments on
the manuscript.}

\bibliographystyle{apj}
\bibliography{bib}
\clearpage

\begin{deluxetable}{llll}
\footnotesize
\tablecaption{Radio- and Far-Infrared-Wavelength
Properties of Sgr D 7 and D 8\label{d7d8}}
\tablewidth{0pt}
\tablehead{
\colhead{Radio name} &
\colhead{Sgr D 7} &
\colhead{Sgr D 8}&
\colhead{References}
}

\startdata
$\alpha_{B19500}$      & 17 45 01.3           & 17 45 47.1          &1,2 \\
$\delta_{B1950}$       & $-$27 42 17          & $-$27 43 45         & 1,2 \\
$l$                    &1.330                 &1.397                &1,2 \\
$b$                    &0.051                 & $-$0.006            &  1,2 \\
$S$(1.4 GHz)(mJy)      & --                   & 26                  &1\\
$S$(1.6 GHz)(mJy)      & 222                  & 21                  &2\\
$S$(5 GHz)(mJy)        & --                   &  24.2               &1\\
IRAS Name              & 17450$-$2742         & 17457$-$2743        &3\\
$S(12 \mu{\rm m})$(Jy) & 9.79:                & 5.54L               &3\\
$S(25 \mu{\rm m})$(Jy) & 28.30                & 3.13                &  3\\
$S(60 \mu{\rm m})$(Jy) & 462.40:              & 85.82:              &3\\
$S(100\mu{\rm m})$(Jy) & 782.90:              & 386.60L             &3\\
$T_{\rm D,25-60}$(K)   &47                    &  43.5               & 4\\
$T_{\rm D,25-100}$(K)  &43.5                  & 38.5                & 4\\
$L_{25-60}$(\Lsun)     &  $2.6\times 10^{4}$  & $4.5 \times 10^{3}$ & 4\\
$L_{25-100}$(\Lsun)    & $4.1\times 10^{4}$   & $1.0 \times 10^{4}$ & 4 \\
$M_{25-60}$(\Msun)     & 33                   & 9                   & 4\\
$M_{25-100}$(\Msun)    & 84                   & 43                  & 4\\
\enddata

\tablecomments{\footnotesize Flux densities are denoted by an $S$. $T_{\rm D}$,
$L$, and $M$ are dust temperature, bolometric luminosity, and total mass
determined from the 25 and 60 $\mu$m IRAS data when thus indexes, or from the
25, 60, and 100 $\mu$m data. (1) Becker et al.(1994) (2) Liszt (1992) (3)
SIMBAD, a : denotes an uncertain value and an L an upper limit. (4) Lis 2006,
pers. comm.}

\end{deluxetable}

\begin{deluxetable}{lcrr}
\footnotesize
\tablecaption{Observed Spectral Lines \label{lines}}
\tablewidth{0pt}
\tablehead{
\colhead{Species}&
\colhead{Transition} &
\colhead{Frequency$^{\rm a}$} &
\colhead{$E_{\rm l}^{\rm b}$}\\
&&\colhead{(MHz)} & \colhead{(K)}
}

\startdata

CS   & $2-1$           & 97980.95& 2.4\\
SiO  & $2-1$           & 86848.96& 2.1\\
\meth&\EFGH            & 84521.21&28.4\\
\    &$2_{-1}-1_{-1}E$ & 96739.39& 0.0\\
\    &$2_0-1_0A^+$     & 96741.42& 2.3\\
\    &$2_0-1_0E$       & 96744.58& 7.6\\
\    &$2_1-1_1E$       & 96755.51&15.5\\
\    &$0_0-1_{-1}E$    &108893.94& 0.0\\
\tablecomments{
$^{\rm a}$ Frequencies and lower state energies are taken
from the JPL catalog (http//spec.jpl.nasa.gov/).
$^{\rm b}$  For the \meth\ lines, lower level energies
are relative to the $0_0$ state for $A$-type lines
and relative to the $1_{-1}$ state for E-type lines.
}
\enddata
\end{deluxetable}

\clearpage
\begin{deluxetable}{rrrrr}
\footnotesize
\tablecaption{Fiducial Positions \label{positions}}
\tablewidth{0pt}
\tablehead{
\colhead{Nr.}& \colhead{$l_{\rm II}$} & \colhead{$b_{\rm II}$} &
\colhead{$\alpha_{{\rm J}2000}$} & \colhead{$\delta_{{\rm J}2000}$}}

\startdata

 1 & 1{\rlap.}{$^\circ$}5750 & $-$0{\rlap.}{$^\circ$}0183 &
17$^{\rm h}$ 49$^{\rm m}$ 23.7 & $-$27$^\circ$ 35' 53''\\
 2 & 1.5939 & $-$0.0148 &  17 49 18.6 & $-27$ 33 54\\
 3 & 1.6418 & $-$0.0641 &  17 49 43.6 & $-27$ 33 52\\
 4 & 1.2829 & $-$0.0289 &  17 48 31.9 & $-27$ 49 27\\
 5 & 1.3690 & $-$0.1000 &  17 49 13.9 & $-27$ 49 04\\
 6 & 1.3168 & $-$0.0650 &  17 49 00.8 & $-27$ 50 55\\
 7 & 1.4110 &   +0.0260 &  17 48 51.7 & $-27$ 42 29\\
 8 & 1.8053 & $-$0.3391 &  17 48 01.6 & $-27$ 33 08\\
 9 & 1.48211&   +0.0264 &  17 45 51.6 & $-27$ 38 22\\ 
\enddata

\tablecomments{
The first column gives the numbers of the fiducial
positions as used in Fig \ref{fig:channelmaps} and Table \ref{gauss}.
The remaining columns give galactic
coordinates and J2000 equatorial coordinates.}

\end{deluxetable}

\begin{deluxetable}{rlrllll}
\footnotesize
\tablecaption{Gaussian Fitting Results \label{gauss}}
\tablewidth{0pt}
\tablehead{
\colhead{Pos.} &
\colhead{Species}&
\colhead{Transition} &
\colhead{$T_{\rm A}^*$} &
\colhead{$\int T_{\rm A}^* dv$} &
\colhead{$v_{\rm LSR}$} &
\colhead{$\Delta v$}\\
&&& \colhead{(K)} & \colhead{(K km$^{-1}$)} &
\colhead{(km s$^{-1}$)} & \colhead{(km s$^{-1}$)}
}

\startdata

 1 &  \meth & \EFGH & \pz 0.53 &\pz 4.9(0.3) &\pz 56.2(0.2) & \pz 8.7(0.6) \\
   &        & \ohn\ & $<0.57$  & --       & --           & --  \\
   &   CS   & $2-1$ & \pz 0.31 &\pz 4.6(4.1) &\pz 48.0(5.5) &    13.8(6.4) \\
   &        &       & \pz 0.71 &   11.0(4.5) &\pz 61.2(2.3) &    14.4(3.3) \\
   &        &       & \pz 0.28 &\pz 8.6(1.3) &   163.2(2.1) &    28.5(5.5) \\
   &   SiO  & $2-1$ & \pz 0.23 &\pz 6.9(0.4) &\pz 53.4(0.6) &    27.4(1.1) \\
   &        &       & \pz 0.12 &\pz 1.2(0.3) &\pz 60.1(0.6) & \pz 9.9(1.4) \\
   &        &       & \pz 0.10 &\pz 4.1(0.2) &   163.3(1.0) &    37.0(2.3) \\
 2 &  \meth & \EFGH & \pz 0.20 &\pz 3.1(0.3) &\pz 51.6(0.8) &    14.1(1.7) \\
   &        &\zkok\ & \multicolumn{4}{c} {fitting impossible}              \\
   &        &\zkok\ & \multicolumn{4}{c} {fitting impossible}              \\
   &        &\ohn\  & $<0.57$  & --       & --           & --              \\
   &   CS   & $2-1$ & \pz 0.23 &\pz 1.5(0.9) &\pz 43.7(1.1) & \pz 5.8(3.2) \\
   &        &       & \pz 0.62 &   12.2(1.2) &\pz 58.7(0.9) &    18.6(2.2) \\
   &        &       & \pz 0.46 &   12.4(1.0) &   162.3(1.0) &    25.1(2.2) \\
   &  SiO   & $2-1$ & \pz 0.17 &\pz 1.4(0.3) &\pz 42.5(0.4) & \pz 7.7(1.3) \\
   &        &       & \pz 0.37 &   10.1(0.5) &\pz 60.3(0.6) &    26.0(1.5) \\
   &        &       & \pz 0.21 &\pz 7.0(0.3) &\  157.9(0.6) &    30.8(1.6) \\
 3 &  \meth & \EFGH & \pz 0.44 &\pz 4.2(0.2) &\pz 51.9(0.3) & \pz 8.9(0.8) \\
   &        &\zkok  & \pz 1.3  &   12.6(0.5) &\pz 56.2(0.2) & \pz 9*       \\
   &        &       & \pz 1.4  &   13.0(0.6) &\pz \pz *     & \pz 9*       \\
   &        &       & \pz 1.4  &   13.3(0.4) &\pz \pz *     & \pz 9*       \\
   &        &       & \pz 0.08 &\pz 0.7(0.4) &\pz \pz *     & \pz 9*       \\
   &        &\ohn\  & $<0.57$  & --       & --           & --  \\
   &  CS    & $2-1$ & \pz 0.33 &   13(2)     &\pz $-2(6)$   &    73(15)    \\
   &        &       & \pz 0.61 &   15(2)     &\pz  58(1)    &    23(2)     \\
   &        &       & \pz 0.55 &   16(3)     &    162(2)    &    27(4)     \\
   &        &       & \pz 0.15 &\pz 4(3)     &    197(7)    &    27(22)    \\
   &  SiO   & $2-1$ & \pz 0.29 &\pz 4.5(0.7) &\pz  45.4(1.1)&    14(2)     \\
   &        &       & \pz 0.40 &\pz 4.2(0.7) &\pz  56.8(0.4)&\pz  9(1)     \\
   &        &       & \pz 0.06 &\pz 0.6(0.1) &\pz 182.3(1.0)&    11(2)     \\
 4 &  \meth & \EFGH & \pz 0.21 &\pz 8.6(0.5) &    118.5(1.2)&    39.6(2.8) \\
   &        &\zkok\ & \multicolumn{4}{c} {fitting impossible}              \\
   &        &  \ohn & $<0.3$   & --          & --           & --           \\
   &  CS    & $2-1$ & \pz 0.74 &   45(9)     &    106(5)    &    57(7)     \\
   &        &       & \pz 0.20 &   13(9)     &    166(21)   &    61(27)    \\
   &  SiO   & $2-1$ & \pz 0.31 &   20(1)     &    116(1)    &    61 (2)    \\
   &        &       & \pz 0.05 &\pz 1.1(0.2) &    179(2)    &    20(4)     \\
 5 & \meth  &\EFGH  & \pz 0.20 &   11.5(0.7) &\pz  79(8)    &    55(4)     \\
   &  CS    & $2-1$ & \pz 0.25 &\pz 6.5(1.1) &   $-22(3)$   &    24(6)     \\
   &        &       & \pz 0.89 &   33(3)     &\pz  72(2)    &    35(3)     \\
   &        &       & \pz 0.53 &   17(3)     &    109(3)    &    30(5)     \\
   &  SiO   & $2-1$ & \pz 0.32 &   13(2)     &\pz  63(2)    &    39(3)     \\
   &        &       & \pz 0.24 &   11(2)     &    101(3)    &    42(4)     \\
 6 & \meth  & \EFGH & \pz 0.24 &\pz 0.5(0.1) &\pz  17.9(0.3)&\pz  1.8(0.7) \\
   &        &       & \pz 0.22 &\pz 5.5(0.4) &\pz  80.8(1.0)&    24.1(2.1) \\
   & CS     & $2-1$ & \pz 0.42 &   38(4)     &\pz  48(4)    &    84(7)     \\
   &        &       & \pz 0.82 &   25(3)     &\pz  84(1)    &    29(3)     \\
   & SiO    & $2-1$ & \pz 0.15 &\pz 5(1)     &\pz  44(2)    &    31(3)     \\
   &        &       & \pz 0.39 &   14(1)     &\pz  83(1)    &    33(1)     \\
 7 & \meth  & \EFGH & $<0.2$   & --          & --           & --           \\
   & CS     & $2-1$ & \pz 0.32 &\pz 9(3)     &\pz  82(3)    &    27(5)     \\
   &        &       & \pz 0.32 &   24(4)     &    100(5)    &    69(12)    \\
   & SiO    & $2-1$ & \pz 0.12 &\pz 3.7(0.3) &\pz  72(1)    &    29(2)     \\
   &        &       & \pz 0.15 &\pz 8.6(0.1) &\pz  98(1)    &    54(2)     \\
 8 &  \meth & \EFGH & \pz 0.23 &\pz 8.2(0.6) &\pz  79.0(1.3)&    33.6(3.0) \\
   &  SiO   & $2-1$ & \pz 0.24 &\pz 8(4)     &\pz  71(1)    &    32(6)     \\
   &        &       & \pz 0.16 &   10(4)     &\pz  85(7)    &    60(5)     \\
 9 & \meth  & \zkok & \multicolumn{4}{c} {fitting impossible}              \\
   & \meth  & \EFGH & \pz 0.21 &\pz 5(1)     &\pz  66(1)    &    24(3)     \\
   &        &       & \pz 0.12 &\pz 6(1)     &\pz  99(6)    &    49(8)     \\
\enddata

\tablecomments{Columns are, right to left, position at
which spectrum was taken (from Table \ref{positions}),
species, transition, corrected antenna temperature, integrated corrected
antenna temperature, LSR velocity, and linewidth (FWHM),
with the latter four quantities determined from
multi-component Gaussian fits. Meaningful fitting was in most cases
impossible
for the \zkok\ quartet of methanol and modeling
of these
and other methanol lines are discussed in \S \ref{metstat}.
For the \zkok\ fit results for position 4  $v_{\rm LSR}$ and $\Delta v$
are given for the $\sim 55$ \kms\ system only, with the linewidth
fixed to the value of the \EFGH\ emission at that velocity.
The fitted  velocity is that of the $2_x-1_x A^+$ lines and the
velocities of the other lines were fixed to it.
For that position, the \meth\ equivalents of the CS 162 and 197 \kms\
components were too difficult to fit.
Upper limits given for the \meth, \ohn\ line are 3 times the
$1\sigma$ rms noise.
}
\end{deluxetable}

\begin{deluxetable}{ccccc}
\tablecaption{CH$_3$OH model results: physical parameters\label{para}}
\tablewidth{0pt}

\tablehead{
\colhead{Pos.} &
\colhead{$T_{\rm K}$} &
\colhead{$n($H$_2)$}&
\colhead{$N(C$H$_3$OH-$A$)} &
\colhead{$N(C$H$_3$OH-$E$)} \\
\colhead{} &
\colhead{(K)} &
\colhead{(cm$^{-3})$} &
\colhead{(cm$^{-2})$} &
\colhead{(cm$^{-2})$}
}

\startdata
2&\multicolumn{4}{c}{high velocity clump}\\
&190&3.6$\times$10$^{4}$ &5$\times$10$^{14}$ &5$\times$10$^{14}$\\
&\multicolumn{4}{c}{extended cloud}\\
&30&8$\times$10$^{4}$ &9$\times$10$^{14}$ &9$\times$10$^{14}$\\
\hline
3&\multicolumn{4}{c}{high velocity clump}\\
&16&4 $\times$10$^4$&1$\times$10$^{14}$ &1$\times$10$^{14}$\\
&\multicolumn{4}{c}{extended cloud}\\
&60&6$\times$10$^4$&6$\times$10$^{14}$ &6$\times$10$^{14}$\\
\enddata

\tablecomments{\footnotesize Pos. denotes the position (from Table \ref{positions})
 toward which the fitted spectra were taken. $T_{\rm K}$ and $n($H$_2)$ are the best
 fit values for the kinetic temperature and the molecular hydrogen density, respectively.
$N(C$H$_3$OH-$A)$ and $N(C$H$_3$OH-$E)$ are the best fit column densities of $A$- and $E$-type methanol, respectively. The calculations assume that both the
high velocity clump and the extended cloud  are extended relative to the telescope beam.}
\end{deluxetable}

\begin{deluxetable}{lrccccc}
\tablecaption{CH$_3$OH model results towards Position 1\label{modeltable}}
\tablehead{
\colhead{Transition}&
\colhead{Frequency}&
\colhead{$T_{obs}$}&
\colhead{$T_{mod}$}&
\colhead{$T_{obs}$}&
\colhead{$T_{mod}$}&
\colhead{References}
\\
&\colhead{(GHz)}&
\colhead{(K)}&
\colhead{(K)}&
\colhead{(K)}&
\colhead{(K)}}
\startdata
&&\multicolumn{2}{c}{extended cloud}&\multicolumn{2}{c}{high velocity clump}\\
2$_0 - 3_{-1}E$&    12.179&  $-$1.25      &$-$0.25   & $-$0.38      &$-$0.24&1\\
4$_{-1} - 3_{0}E$&  36.169&   $\sim 0.4$   &0.5       &$ \sim 0.4$ &0.26&2\\
5$_{-1} - 4_{0}E$&  84.521&0.34&0.23&0.26&0.10&3\\
2$_{-1}\to 1_{-1}E$&96.739&&&&&\\
2$_{0}\to 1_{0}A$  &96.741$^a$&1.85&1.63&1.05&0.66&3\\
2$_{0}\to 1_{0}E$  &96.745&&&&&\\
0$_0 - 1_{-1}E$&   108.894&   not detected$^b$  &0.2       & not detected$^b$    &0.08&3\\
8$_0 - 8_{-1}E$&   156.489&   0.00         &0         &0.00        &0&2\\
2$_1 - 3_{0}A$&    156.602&   $-$0.17      &$-$0.08   &$-$0.13     &$-$0.05&2\\
7$_0 - 7_{-1}E$&   156.829&   0.00         &0         &0.00        &0&2\\
6$_0 - 6_{-1}E$&   157.049&   0.00         &0         &0.01        &$-$0.003&2\\
5$_0 - 5_{-1}E$&   157.179&   0.00         &0         &0.00        &$-$0.005&2\\
4$_0 - 4_{-1}E$&   157.246&   0.00         &0         &0.00        &$-$0.01&2\\
1$_0 - 1_{-1}E$&   157.271&                &       &            &\\
3$_0 - 3_{-1}E$&   157.272$^c$&  0.09      &0.2             &0.17 &0.11&2\\
2$_0 - 2_{-1}E$&   157.276&                &       &            &\\
\enddata

\tablecomments{\footnotesize $T_{obs}$ and $T_{mod}$ are the
observed and the modeled antenna temperatures, respectively.
$^a$Blend of 2$_k - 1_k,k=0,\pm1$-$A,E$ lines.
$^b$with an rms of 0.2~K.
$^c$Blend of 1$_0 - 1_{-1}E$, 3$_0 - 3_{-1} E$ and 2$_0 - 2_{-1} E$ lines. (1) \citet{WhiteoakPeng1989}; (2) \citet{Salii_etal2002}; (3) this work.}
\end{deluxetable}

\clearpage

\begin{figure}[htb]
  \centering
    \includegraphics[width=18cm]{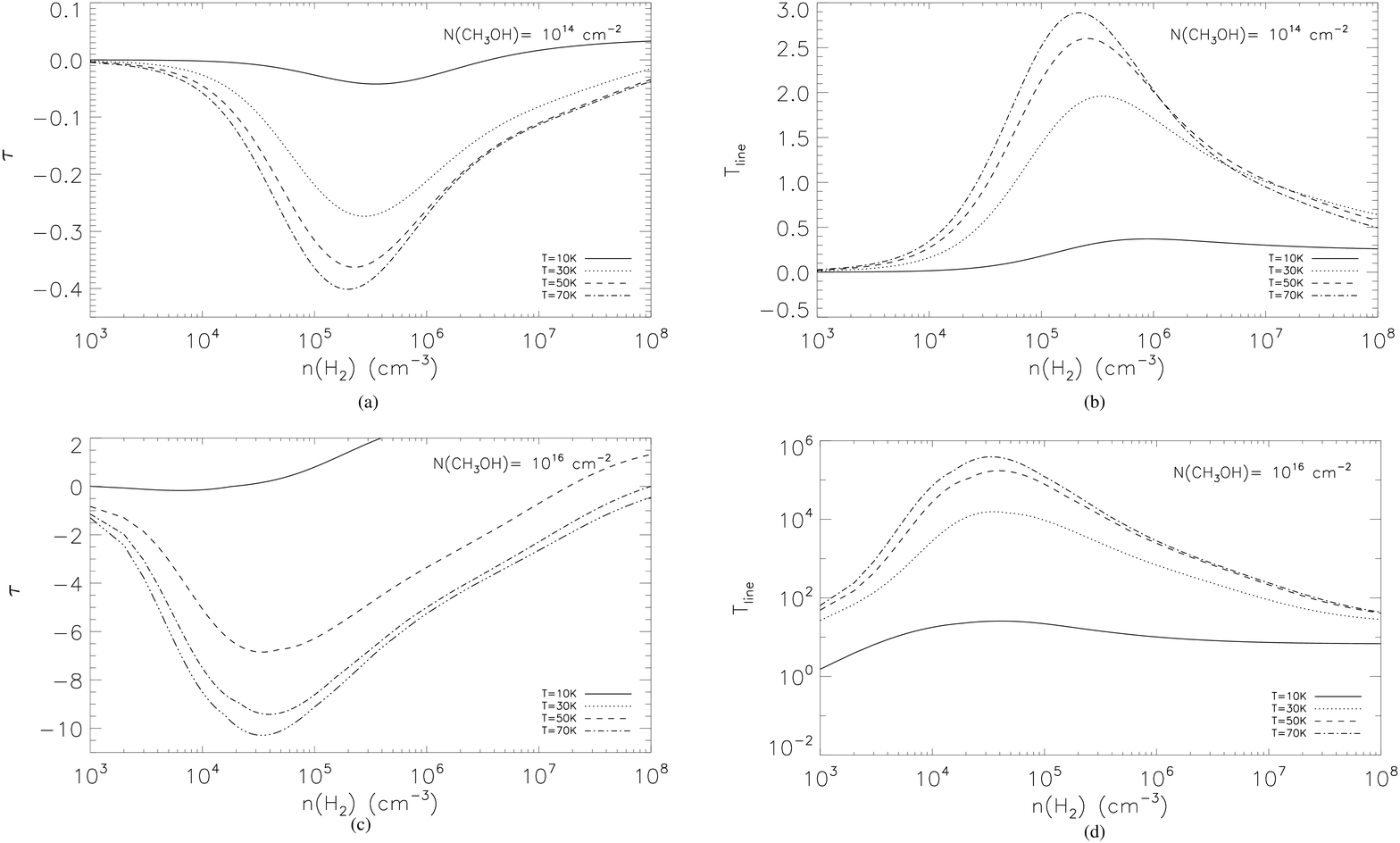}
\caption{Line optical depths (left panels) and line intensities (right panels) of 5$_{-1} - 4_{0} E$ line
as function of different temperatures, for two  CH$_3$OH-$A$ column densities, $10^{14}$ cm$^{-2}$ (upper panels)
and $10^{16}$ cm$^{-2}$ (lower panels).}\label{tau84}
\end{figure}
\begin{figure}[htb]
  \centering
    \includegraphics[width=9cm]{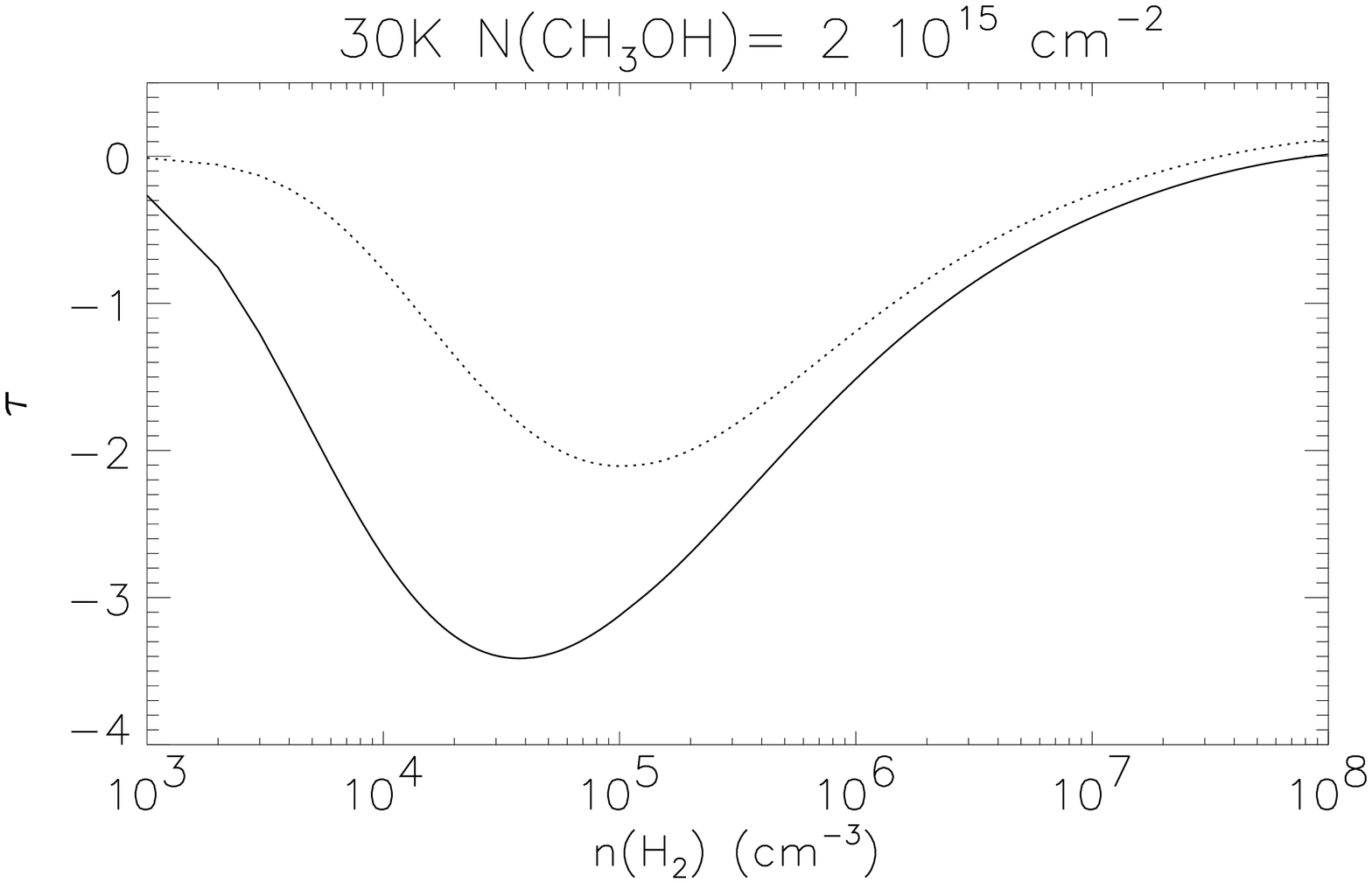}
    \includegraphics[width=9cm]{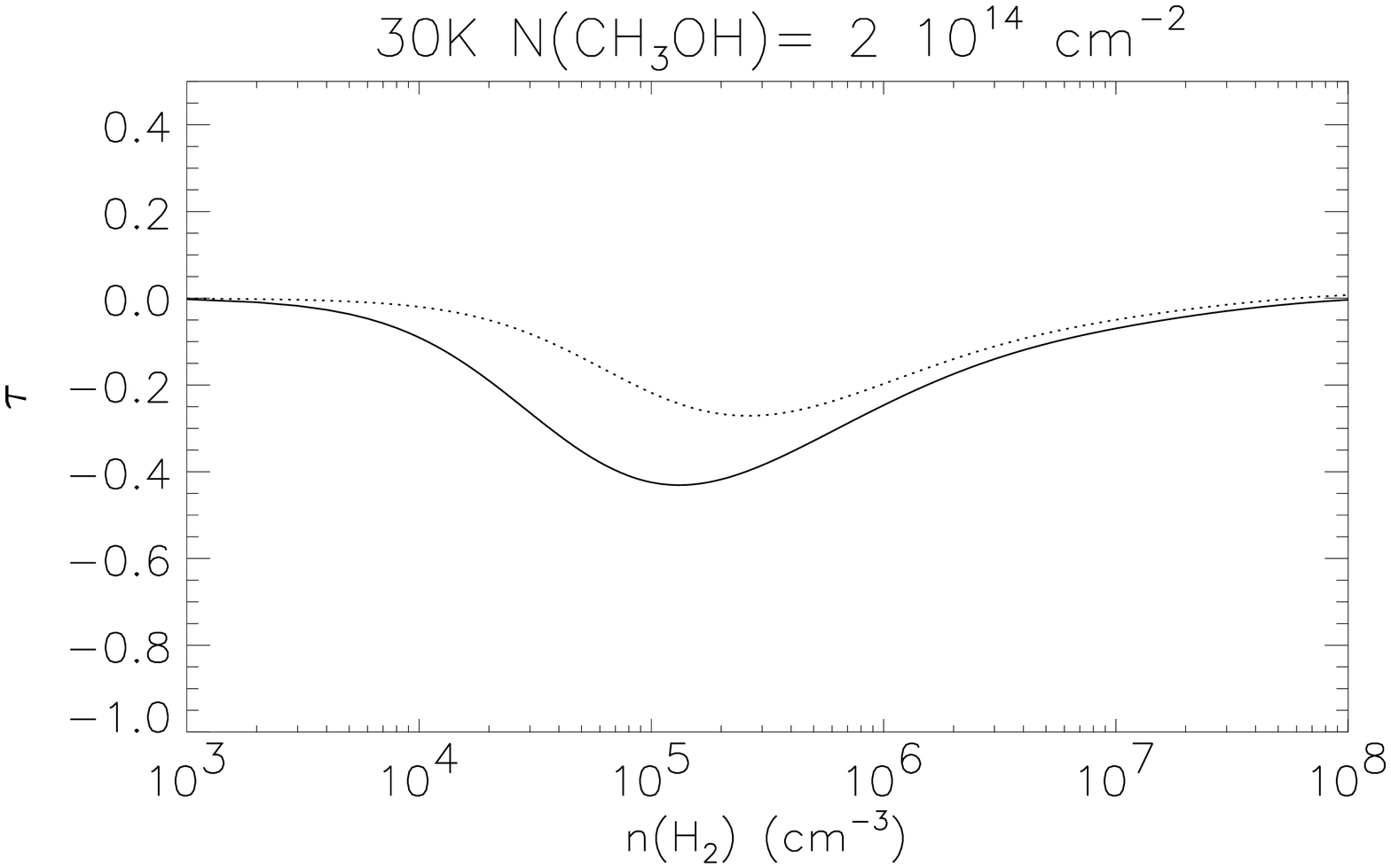}
\caption{Line optical depths of 4$_{-1} - 3_{0} E$ \textit{(solid line)}
and  5$_{-1} - 4_{0} E$ \textit{(dotted line)}
as function of density at different  CH$_3$OH column densities, $2~10^{15}$ cm$^{-2}$ \textit{upper panel} and $2~10^{14}$ cm$^{-2}$ \textit{lower panel}.}\label{tau}
\end{figure}


\end{document}